\documentclass[]{aa}
\RequirePackage{amsmath}
\RequirePackage{amssymb}        
\RequirePackage{epsfig}		
\usepackage{graphicx,epsfig,fancyhdr,psfig,rotating,amsmath,epsf,txfonts,natbib}
\usepackage[font=sf]{caption}

\def\etal{{et al. }}

\def\3o{O~{\sc iii}}
\def\4o{O~{\sc iv}}

\def\arcsec{$^{\prime\prime}$}
\newcommand{\comment}[1]{}

\begin{document}

\title{What is the true nature of blinkers?}
\author{S. Subramanian\inst{1},  M. S. Madjarska\inst{1}, J. G. Doyle\inst{1} and D. Bewsher\inst{2}}
\offprints{M.S. Madjarska, madj@arm.ac.uk}
\institute{Armagh Observatory, College Hill, Armagh BT61 9DG, N. Ireland
\and Jeremiah Horrocks Institute, University of Central Lancashire, Preston, Lancashire, PR1 2HE, UK} 
\date{Received date, accepted date}
%%-----------------------------------------------------------------------
\abstract
{}
{The aim of this work is to identify the true nature of the transient EUV brightenings, 
called blinkers.}
{Co-spatial and co-temporal multi-instrument data, including imaging (EUVI/STEREO, XRT and SOT/Hinode), 
spectroscopic (CDS/SoHO and EIS/Hinode) and magnetogram (SOT/Hinode) data, of an isolated 
equatorial coronal hole were used. An automatic program for identifying transient brightenings 
in CDS O~{\sc v}~629~\AA\ and EUVI~171~\AA\  was applied.}
{We identified 28 blinker groups in the CDS O~{\sc v}~629~\AA\ raster images.  All CDS O~{\sc v}~629~\AA\ blinkers showed counterparts in EUVI 171~\AA\ and  304~\AA\  images. We classified these blinkers into two categories, one associated with coronal counterparts and other with no coronal counterparts as seen in XRT images and EIS~Fe~{\sc xii}~195.12~\AA\ raster images. Around two-thirds of the blinkers  show coronal counterparts and correspond to various events like EUV/X-ray jets, brightenings in coronal bright points or foot-point brightenings of larger loops. These brightenings occur repetitively  and have a lifetime of around 40 min at transition region temperatures. The remaining blinker groups with no coronal counterpart in XRT and EIS~Fe~{\sc xii}~195.12~\AA\ appear as  point-like brightenings and have chromospheric/transition region origin. They take place  only once and have a lifetime of around 20 minutes. In general, lifetimes of blinkers are different at different wavelengths, i.e. different temperatures, decreasing  from the chromosphere to the corona. }
{This work shows that the term blinker covers a range of phenomena. Blinkers are the EUV response 
of various transient events originating at coronal, transition region and chromospheric heights. Hence, events associated 
with blinkers contribute to the formation and maintenance of the temperature gradient in the transition region and the corona.}

\keywords{Sun: atmosphere -- Sun: corona -- Methods:observational -- Methods: data analysis }

\authorrunning{Subramanian \etal}
\titlerunning{The nature of blinkers}

\maketitle
%%-----------------------------------------------------------------------
\section{Introduction}
The quasi-steady mechanism(s) that sustains the heating of the outer solar atmosphere and the 
solar wind  is currently under intensive investigation using state-of-the-art ground- and 
space-based instrumentation. These observations show that small-scale salt and pepper like 
bipolar network magnetic fields constantly reconfigure themselves on time scales of minutes-to-hours, 
resulting in a complicated and dynamically evolving solar atmosphere. It is believed that magnetic reconnection and subsequent energy release is the major mechanism which can relax these constantly evolving magnetic field structures. Such small-scale energy releases seen as sudden brightenings (blinkers or extreme-ultraviolet (EUV) brightenings), fast plasma ejections (spicules or EUV/X-ray jets), etc., are omnipresent and have been reported at different heights in the solar 
atmosphere. It is crucial to establish links between these different events as they potentially 
connect the lower and upper solar atmosphere and, hence, could contribute to the transfer of mass 
and energy in the atmosphere \citep{1988ApJ...330..474P, 1997ApJ...487..424S, 
1998Natur.394..152S, 1999ApJ...526..505M, 2002ApJ...565.1298W, 2005ApJ...629..572Y}.

%%%%%%%%%%% Fig 1%%%%%%%%%%%
\begin{figure*}[htp!]
\begin{center}
\includegraphics[scale=0.5]{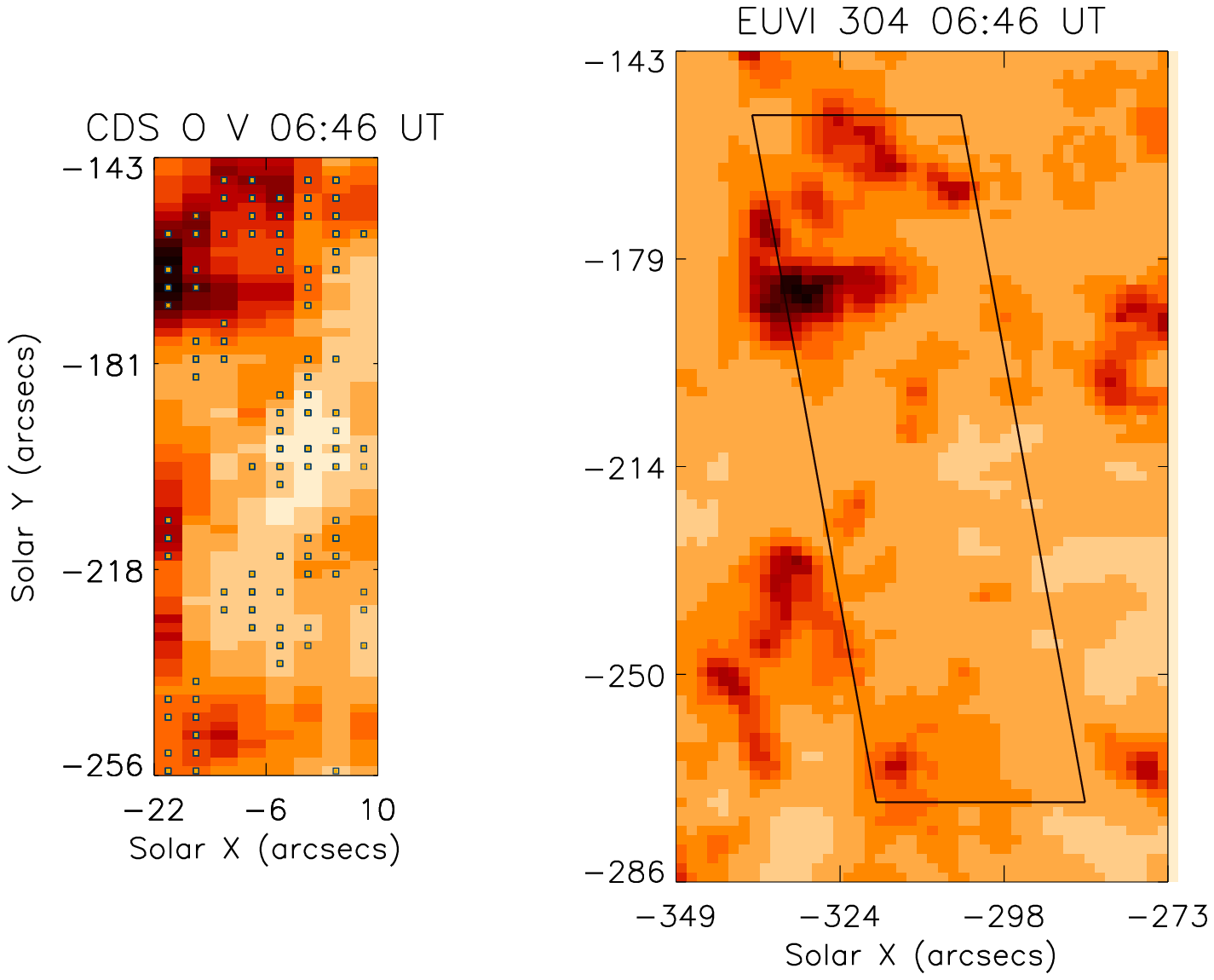}
\hspace{0.07in}
\includegraphics[scale=0.5]{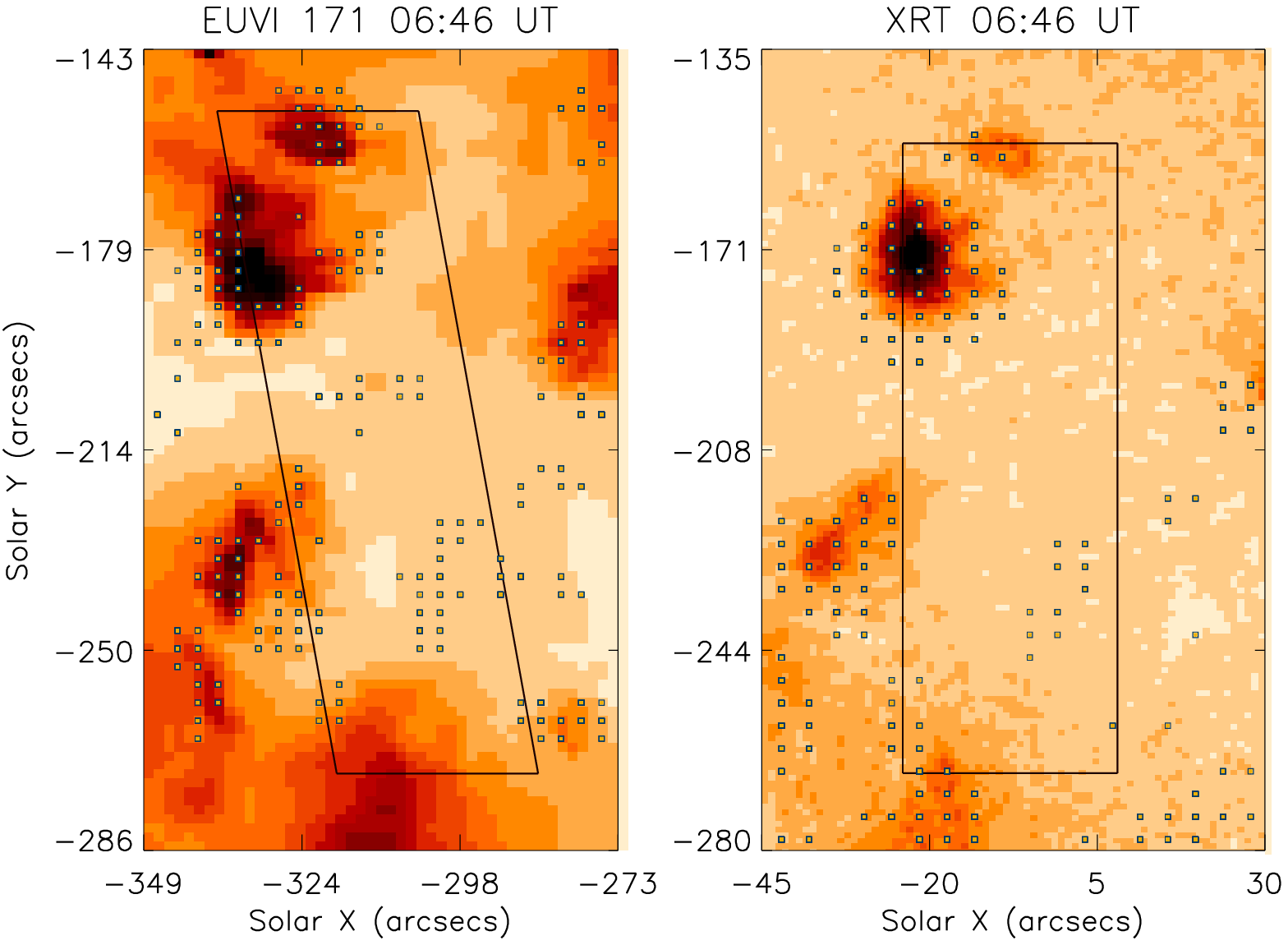}
\vspace{-0.1in}
\caption{ From left: CDS O~{\sc v}~629~\AA\ raster, EUVI He~{\sc ii}~304~\AA\ image, 
EUVI 171~\AA\ image and XRT Al$\_$poly image. The EUVI and XRT images have the 
field of view of the CDS raster over-plotted as a black rectangular box. The identified 
brightenings in CDS, EUVI and XRT are over-plotted with boxes in the respective images. 
The difference in the orientation and the coordinates of the EUVI image with respect to the CDS and XRT field-of-views 
is due to the difference in the viewing angles between STEREO-A and the SoHO and Hinode spacecrafts.}
\vspace{-0.25in}
\label{fig1} %chbr_all_rasters
\end{center}
\end{figure*}

EUV brightenings, also called blinkers, were first reported by \citet{1997SoPh..175..467H} in 
the quiet Sun, at transition region (TR) temperatures, using the Coronal Diagnostic Spectrometer 
\citep[CDS;][]{1995SoPh..162..233H} on-board SoHO spacecraft. They show intensity enhancements of a 
factor of 2--3 \citep{1997SoPh..175..467H,2002SoPh..206...21B} and Doppler velocities of 25--30~km~s$^{-1}$ 
\citep{2003SoPh..215..217B} in TR lines like O~{\sc v}~629.77~\AA\ (T~$\approx$~2.4~$\times$~10$^{5}$~K). The 
average lifetime of these EUV brightenings is 16 minutes ranging from 6 to 40 min \citep{2002SoPh..206...21B}. They 
have also been observed in active regions 
\citep{1997ESASP.404..717W,2002SoPh..206..249P,2003SoPh..215..217B}, with slightly higher Doppler
velocities of 20--40~km~s$^{-1}$ \citep{2003SoPh..215..217B}. Corresponding signatures of blinkers 
were  found in chromospheric lines \citep{2001A&A...373.1056B,2003A&A...406..363B,2004ApJ...602.1051B,2004ApJ...611.1125B}. 
\citet{1999A&A...351.1115H} derived an average intensity increase of 4\% and 7\% in the coronal Mg~{\sc ix} 
368.9~\AA\ and Mg~{\sc x}~624.9~\AA\ lines, respectively, for these events.  \citet{2002SoPh..206...21B}
also detected a weak response of blinkers in these lines  and concluded ``that blinkers
have no coronal signatures'', although ``it may be simply that these lines are too weak to detect anything in''. 
\citet{2002SoPh..205..249P} suggested from simple physical models that blinkers can be produced by         
five different physical
mechanisms, namely: the heating of cool spicular material; the containment of plasma in low-lying
loops in the network; the thermal linking of cool and hot plasma at the feet of coronal loops; the
heating and evaporating of chromospheric plasma in response to a coronal heating event; and the
cooling and draining of hot coronal plasma when coronal heating is switched off. 

\comment{\citet{2008A&A...488..323S} showed that some blinkers are associated with the emergence of magnetic flux and the formation of a new loop structure seen in TRACE~1550~\AA\ images. A 3-dimensional magnetic topology analysis  suggested  a magnetic 
reconnection across a coronal magnetic null point may have been responsible for the loop formation.}

Coronal jets are dynamic features which are observed as collimated ejections of plasma on small 
scales. They were first observed with the solar X-ray telescope onboard the Yohkoh satellite 
\citep{1992PASJ...44L.173S} and are believed to be the result of magnetic reconnection 
\citep{1995Natur.375...42Y}. The X-ray Telescope \citep[XRT;][]{2007SoPh..243...63G} onboard 
Hinode opened a new era for studying X-ray features in tens-of-second detail, revealing the association 
of some jets with the expansion and eruption of coronal bright point loops 
\citep{2007PASJ...59S.745S,2009SoPh..254..259F,2011A&A...526A..19M}. More recently, \citet{2010A&A...516A..50S} completed a statistical study of X-ray brightenings 
in coronal holes and quiet Sun regions. The authors found that more than 70\% of  the 
brightenings observed in coronal holes were coronal jets. However, very few jet-like 
events were observed in the quiet Sun. The authors also suggested that the remaining unresolved 
X-ray brightenings could involve two sided loop  reconnection \citep{1994xspy.conf...29S} 
between the emerging flux and the overlying coronal field or flows in loop structures perhaps 
triggered by reconnection shocks from  neighbouring regions. For more detailed introduction see \citet{2010A&A...516A..50S} and \citet{2011A&A...526A..19M}.

%%%%%%%%% Fig 2%%%%%%%%%%%%
\begin{figure*}[htp!]
\begin{center}
\includegraphics[scale=0.70]{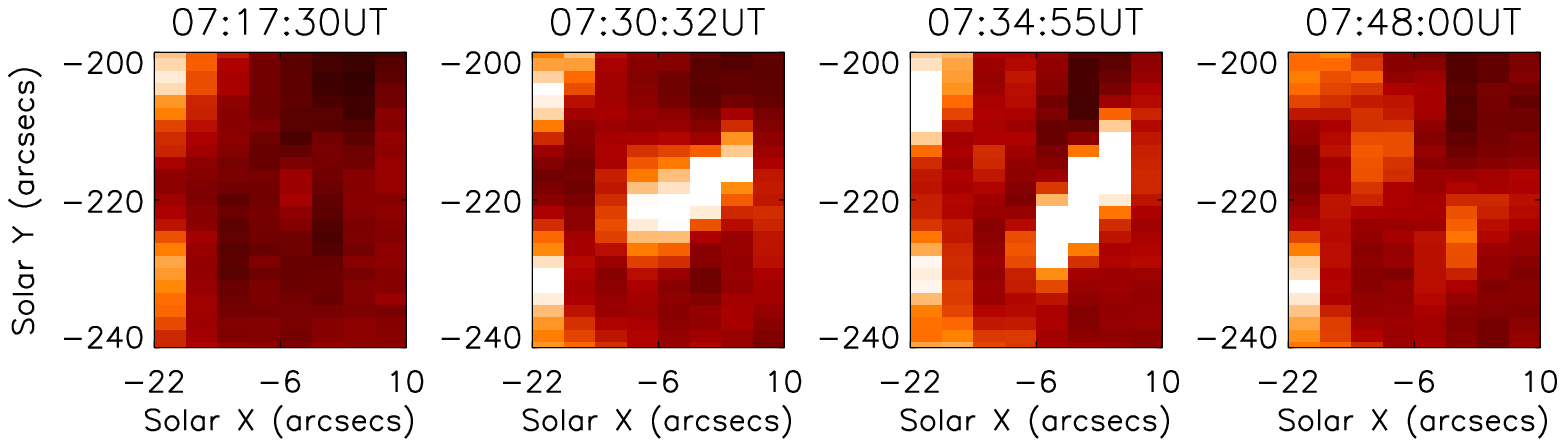}
\includegraphics[scale=0.70]{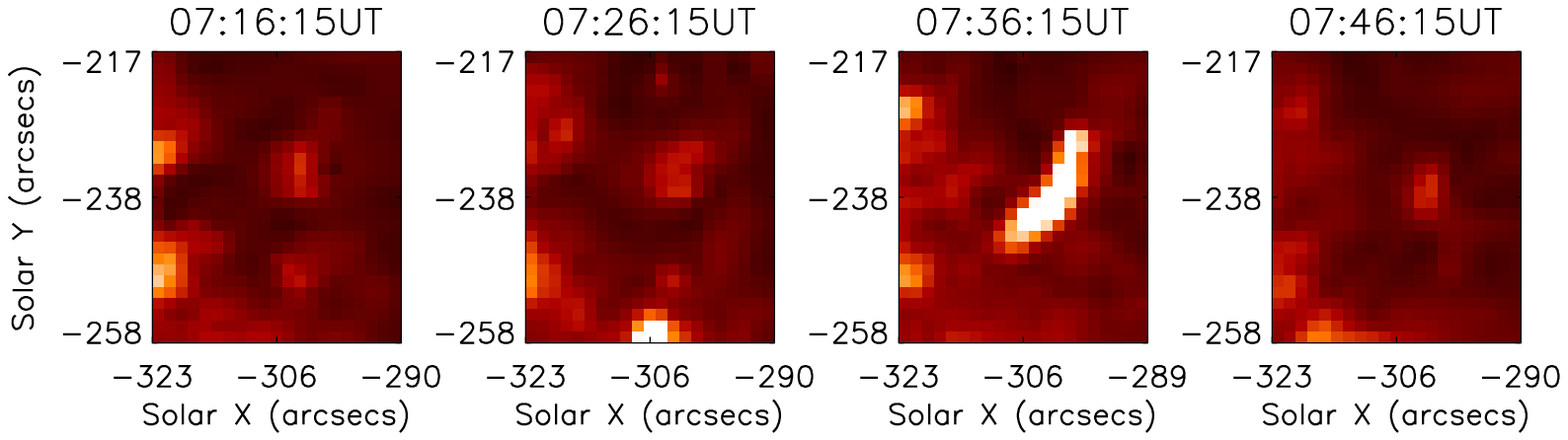}
\includegraphics[scale=0.70]{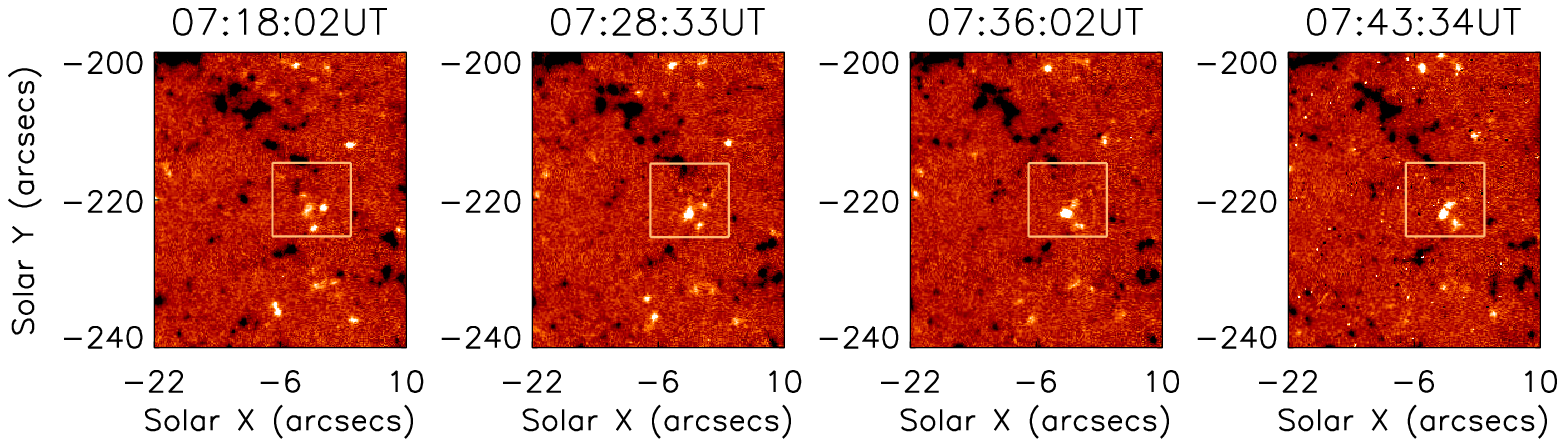}
\includegraphics[scale=0.70]{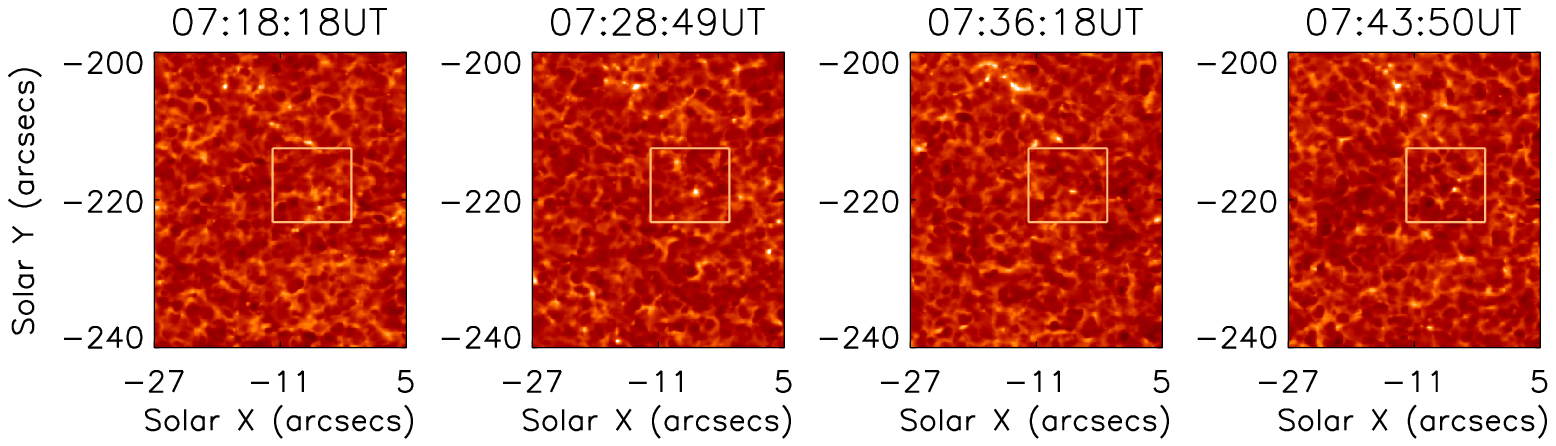}
\includegraphics[scale=0.70]{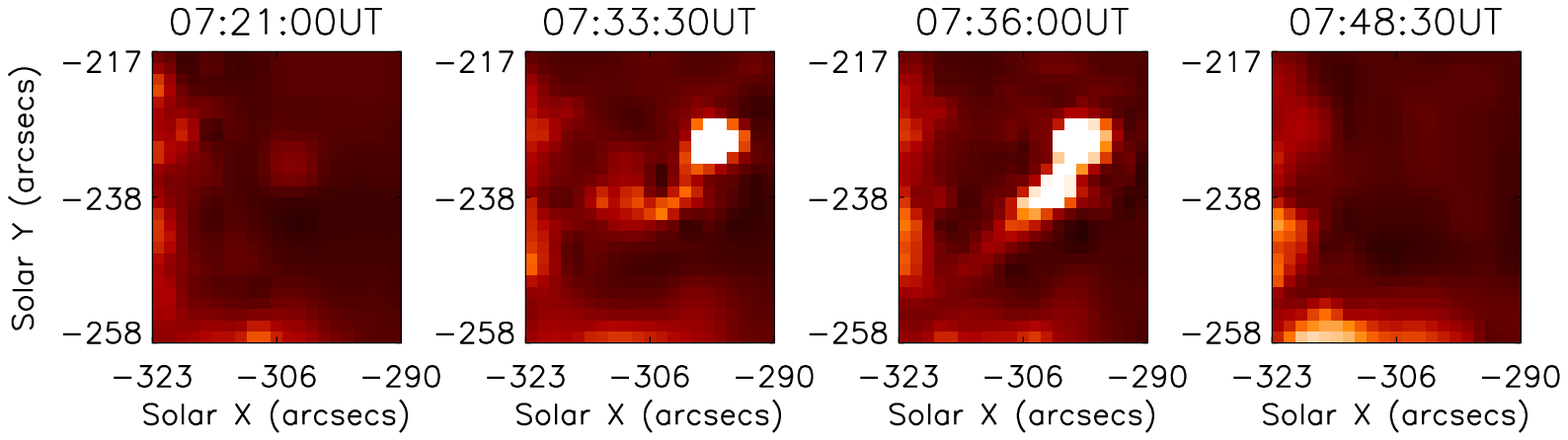}
\includegraphics[scale=0.70]{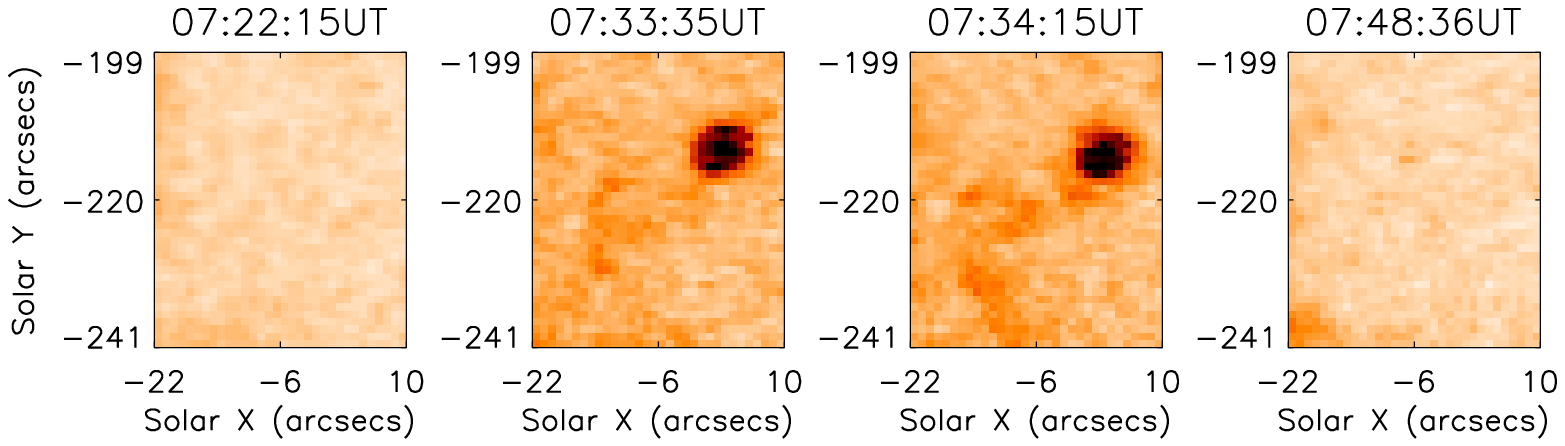}
\caption{ Blinker group 1 from top in CDS O~{\sc v}~629~\AA, EUVI~304~\AA, 
SOT FG magnetograms, SOT Ca~{\sc ii}~H, EUVI~171~\AA\
and XRT Al$\_$poly. The corresponding magnetogram field-of-view from which the lightcurve of the 
positive flux was derived,  is outlined by the white box on the SOT FG magnetograms and Ca~{\sc ii}~H images.}
\label{fig2}%chbr_br13}
\end{center}
\end{figure*}

An intensity increase, i.e. a brightening, either seen in imager or spectrometer data  can be a 
signature of various physical processes. This study aims at investigating the signature of which  
phenomena the EUV brightenings, often referred to as blinkers, are?  We used state-of-art co-observations 
from imagers and spectrometers which presently observe the Sun with unprecedented spatial and temporal 
resolution as well as a spectral coverage of  a temperature range which provides a reliable coronal diagnostic.   

The paper is organized as follow: Section 2 describes the observational material. In Section~3 we outline 
the procedure for blinker identification in images taken in spectral lines or imager passbands covering different temperatures, i.e. chromospheric, transition region or coronal. Section~4 presents the obtained results.  
The discussion and conclusions are given in Section 5.

\section{Data reduction and preparation}
\label{data}

Observations obtained with the Solar Optical Telescope \citep[SOT;][]{2008SoPh..249..167T},
the EUV imaging spectrometer \citep[EIS;][]{2005AdSpR..36.1494C,2007SoPh..243...19C}) and 
XRT all onboard Hinode, the CDS onboard SoHO and the EUV Imager \citep[EUVI;][]{2008SSRv..136...67H} 
onboard STEREO were used in the analysis. Approximately 10 hours of data were taken during a dedicated observing 
run of an isolated equatorial coronal hole (ECH) on 2007 November 12. 

X-ray images from XRT/Hinode were taken with the Al$\_$poly filter with a pixel size of 1\arcsec 
$\times$ 1\arcsec, a field-of-view (FOV) of 370\arcsec $\times$ 370\arcsec\ and a cadence of 40 seconds. The data were reduced using 
\textit{xrt\_prep.pro} which includes despiking, normalisation to data numbers per second 
to account for variations in exposure times, satellite jitter and orbital variation corrections. 

EIS/Hinode was observing in a rastering mode with the 2\arcsec\ slit. The data consist of a 
large raster with a FOV of 120\arcsec $\times$ 496\arcsec\ followed by 34 small rasters with 
a FOV of 24\arcsec $\times$ 248\arcsec\ taken with a cadence of 12.5 minutes, and a pixel 
size of 2\arcsec $\times$ 1\arcsec. Many spectral lines were available for analysis, however, 
in the present study only the O~{\sc v} 192.9~\AA\ (logT$_{\mbox{max}}$ = 5.4~K), 
Fe~{\sc viii}~185.21~\AA\ (logT$_{\mbox{max}}$ = 5.8~K), Fe~{\sc x}~184.59~\AA\ (logT$_{\mbox{max}}$ = 6.0~K) 
and Fe~{\sc xii}~195.12~\AA\ (logT$_{\mbox{max}}$ = 6.1~K) lines are used. The data were reduced 
using \textit{eis\_prep.pro} and were corrected for EIS detector tilt and satellite orbital variation.

SOT/Hinode provided both Ca~{\sc ii} H images and magnetograms. The Ca~{\sc ii} H images were taken 
with a pixel size of 0.11\arcsec $\times$ 0.11\arcsec\ and a cadence of approximately 90 seconds. 
The magnetograms were taken using the Na~{\sc i}~5896~\AA~filter with a pixel size of 
0.16\arcsec $\times$ 0.16\arcsec\ and a cadence of 90 seconds. Both datasets were reduced 
using \textit{fg\_prep.pro}.

Full disk EUV images from EUVI/STEREO-A were taken using the 171~\AA\ filter (dominated mostly by 
Fe~{\sc ix/x}, but in some instances may have a large contribution from TR lines such as 
O~{\sc v} and O~{\sc vi}, more on this below) with a pixel size of 1.6\arcsec $\times$ 
1.6\arcsec\ and a cadence of 150 seconds. The images were reduced with \textit{euvi\_prep.pro}. 

CDS/SoHO observed the He~{\sc i}~584.3~\AA, He~{\sc ii}~303.78~\AA\ and 303.786~\AA\  (seen in second order; hereafter 304~\AA) and 
O~{\sc v}~629.77~\AA\ (hereafter 629~\AA) lines and a spectral window centred around 560~\AA\ containing Ca~{\sc x} 
and Ne~{\sc vi} lines in rastering mode. Only the O~{\sc v}~629~\AA\ data were used in this 
paper. The data consist of a large raster with a FOV of 240\arcsec $\times$ 240\arcsec\ 
followed by 40 small rasters with a FOV of 40\arcsec $\times$ 124\arcsec\ taken with a cadence 
of 4 minutes and a pixel size of 4\arcsec $\times$ 1.68\arcsec. The data were reduced using 
the standard software packages for the correction of missing pixels, cosmic ray hits, CCD bias 
effects and flat-field effects. An intensity map of the emission from the spectral window was 
obtained by integrating over the spectral line. 

The CDS rasters, SOT, EUVI and XRT images were de-rotated to a common reference time in order 
to compensate for the solar rotation. Furthermore, the data from the different instruments were 
co-aligned in order to follow the same region at different temperature regimes. The CDS 
O~{\sc v}~629~\AA, the SOT Ca~{\sc ii}~H and the magnetogram data were co-aligned with respect to the 
TRACE 1550~\AA\ images (which were also taken as part of the dedicated observing run, but 
not used in the scientific analysis presented here). Then the EIS O~{\sc v}~192.90~\AA\ intensity images
were co-aligned with the CDS O~{\sc v}~629~\AA\ ones. The EIS Fe~{\sc xii}~195.12~\AA\ data were 
then cross correlated with the XRT data, which were in turn aligned with the EUVI 171~\AA\ 
data. The coordinates are unique to the EUVI image (Fig.~\ref{fig1}) and cannot be compared with the  
Hinode and SoHO images due to the difference in the viewing angles between STEREO, Hinode and SoHO. 
Figure~\ref{fig1} shows a sample (from left) CDS O~{\sc v}~629~\AA\ raster, EUVI He~{\sc ii}~304~\AA\ 
image, EUVI 171~\AA\ image and XRT image with the field of view of the CDS raster over-plotted 
as a rectangular box. 

For the present study, we used data taken of a coronal hole region as they are ideal for this 
work because of their low background emission at coronal temperatures when compared to the 
quiet Sun or active region corona. The lightcurves were smoothed with a smoothing window of width 5 for 
the Ca~{\sc ii}~H and 3 for the rest of the lightcurves. Them the lightcurve values were normalized with respect 
to the maximum radiance. 

\section{Blinker identification}
\label{identification}
The first step was to identify brightenings in the CDS O~{\sc v}~629~\AA\ rasters. For this we 
applied an automated brightening identification procedure described in \citet{2010A&A...516A..50S}. 
The procedure finds intensity enhancements in the input light-curves which are above a user 
defined intensity threshold together with the corresponding start and end times. The input 
light-curves were made by summing over 1 $\times$ 2 pixels$^{2}$, which is approximately 
4\arcsec\ $\times$ 3.4\arcsec, and were smoothed over a time window of 3 (i.e. over 3 consecutive 
images) to remove unwanted spikiness in the background. As CDS was rastering completely inside 
the coronal hole, with no quiet Sun region, we used a single intensity threshold of 1.45$\times$ 
the background. The program estimates locally the background  for each lightcurve by eliminating 
the identified intensity peaks and averaging over the rest of the lightcurve. A duration threshold 
of 45 minutes was also used as blinkers can have lifetimes from 
6 minutes to around 40 minutes \citep{2002SoPh..206...21B}. 

The same procedure was used to identify brightenings in the EUVI 171~\AA\ images and X-ray images 
with an intensity threshold of 1.35~$\times$ and 2~$\times$ the background, respectively, and a duration 
threshold of 45 minutes in order to identify the counterparts of the CDS blinkers. The input light-curves 
were made by summing over 2 $\times$ 2 pixels$^{2}$, which is approximately 3.2\arcsec\ $\times$ 3.2\arcsec\ 
for the EUVI 171~\AA\ data and 4~$\times$~4 pixels$^2$, which is approximately, 4\arcsec~$\times$~4\arcsec\ 
for the X-ray data. The intensity thresholds used in this work were calculated with a trial and error method.

Figure~\ref{fig1}\comment{chbr_euvi171} shows sample images with the identified 
brightenings over-plotted. A total of 96 brightenings were identified in the CDS O~{\sc v}~629~\AA\ 
raster images. These brightenings were not uniformly distributed over the field of view, instead  
a number of brightenings occurred close to each other, forming clusters. An event bigger 
than the pixel scale of one light-curve will have its imprint in adjacent light-curves. Hence, 
a group of identified brightenings which occurred close together and showed similar light-curves 
and lifetimes, were visually grouped into individual events. We identified 28 such blinker groups over the 
course of our observations in the CDS O~{\sc v}~629~\AA\ data. {In addition to the automatic procedure, each 
brightening identified in CDS O~{\sc v}~629~\AA\ was visually cross-correlated with the EUVI 171~\AA\ and 
X-ray data to search for its counterpart. We found that all CDS O~{\sc v}~629~\AA\ 
blinkers have a counterpart in the EUVI 171~\AA~ images, while only 57\% of them were present in XRT 
images. Thus, we classified the identified blinker groups into two categories, one with an X-ray 
counterpart and the other with no X-ray counterpart.
\section{Results}
\label{results}

%%%%%%%%% Fig 3%%%%%%%%%%%%
\begin{figure}[htp!]
\begin{center}
\includegraphics[scale=1.1]{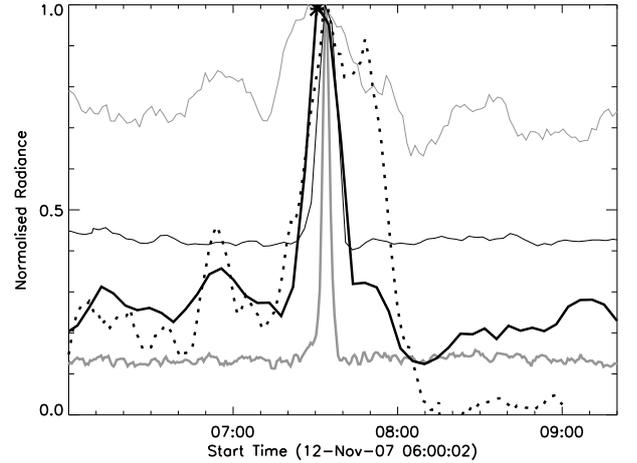}
\caption{Light-curves in CDS O~{\sc v}~629~\AA\ (thick black line), 
EUVI~171~\AA\ (thin black line), SOT Ca~{\sc ii}~H (thin grey line) and XRT Al$\_$poly (thick grey line)
along with the light-curve of the corresponding magnetic fluxes.  The light curve of the positive  
magnetic flux is shown with a dotted line. The asterisk symbol indicates the peak in CDS O~{\sc v}~629~\AA.}
\label{fig3}
\end{center}
\end{figure}

%%%%%%%%% Fig 4%%%%%%%%%%%%

\begin{figure*}[htp!]
\begin{center}
\includegraphics[scale=0.70]{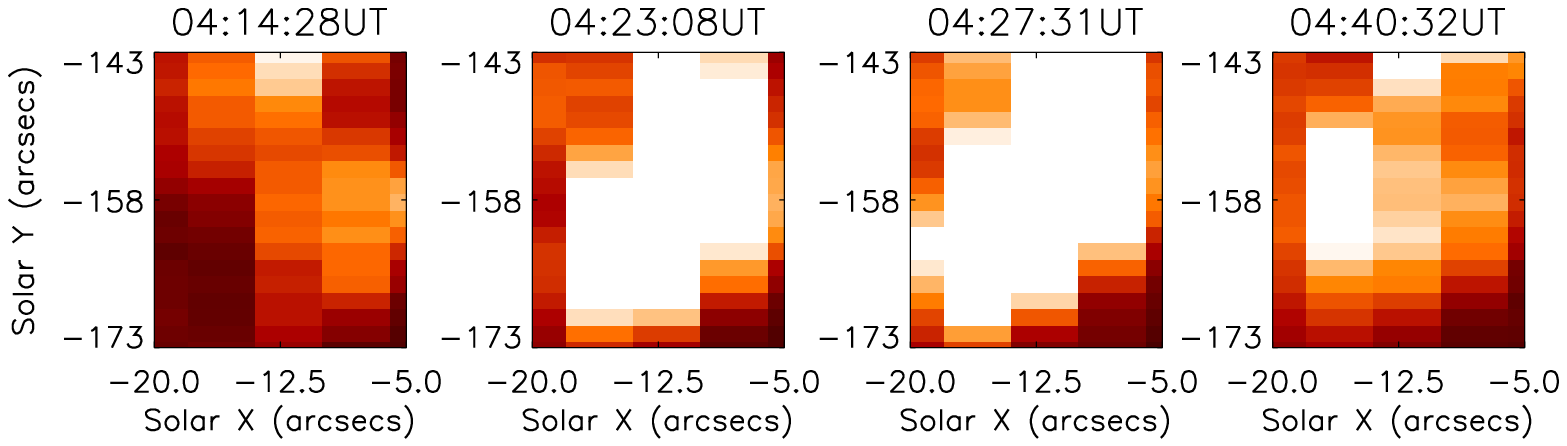}
\includegraphics[scale=0.70]{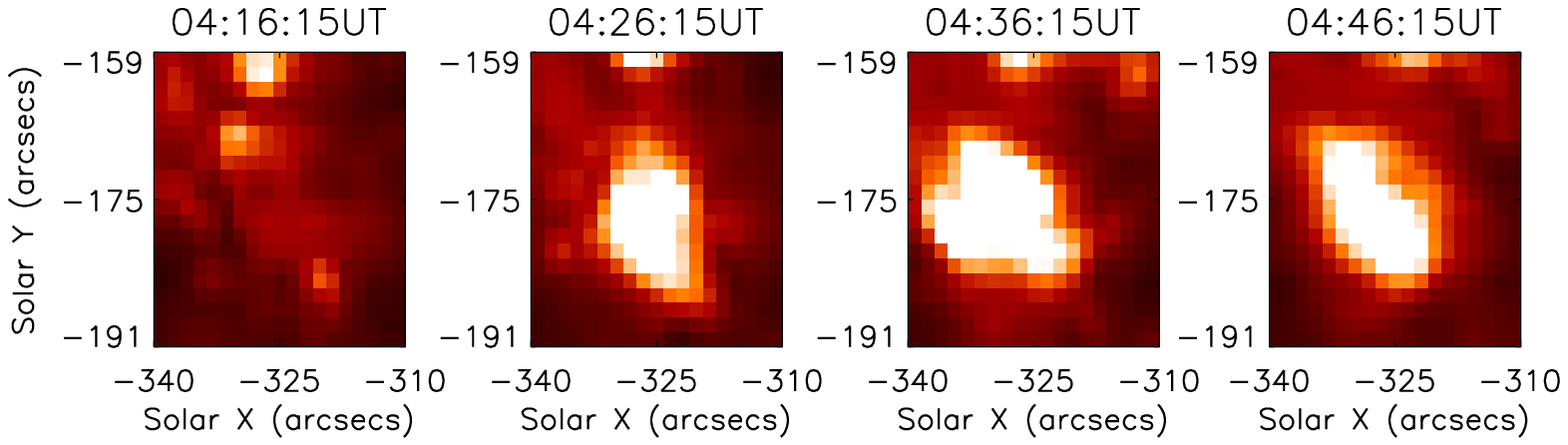}
\includegraphics[scale=0.70]{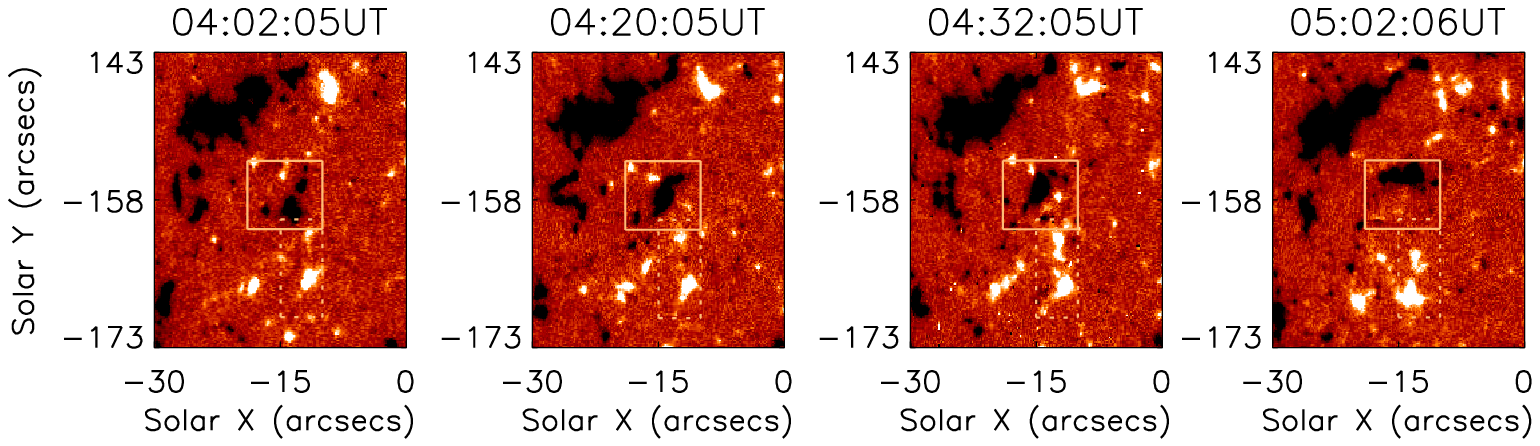}
\includegraphics[scale=0.70]{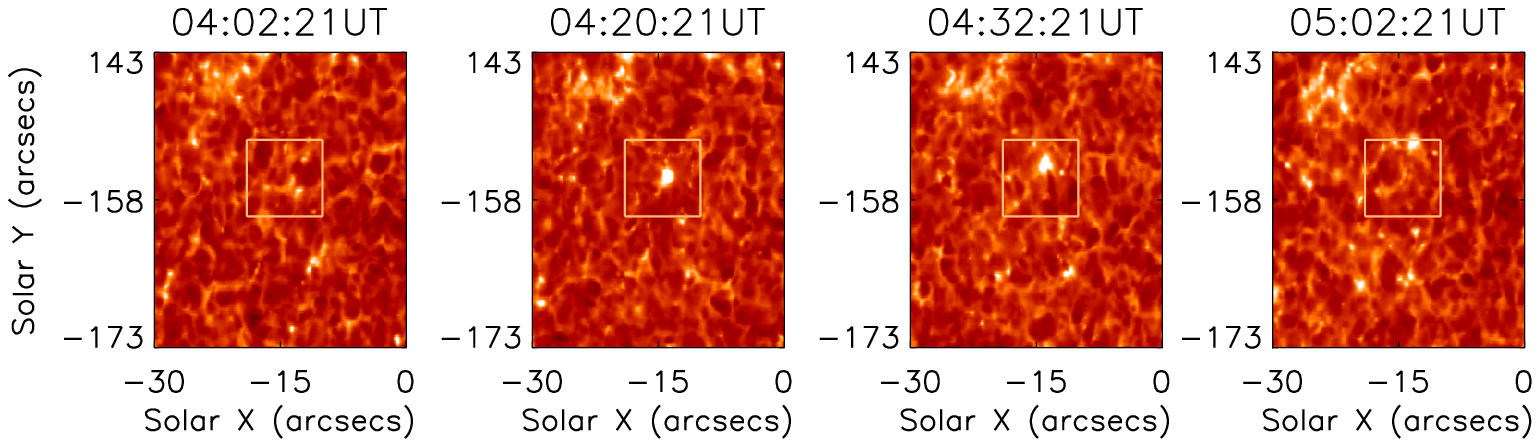}\\
\includegraphics[scale=0.70]{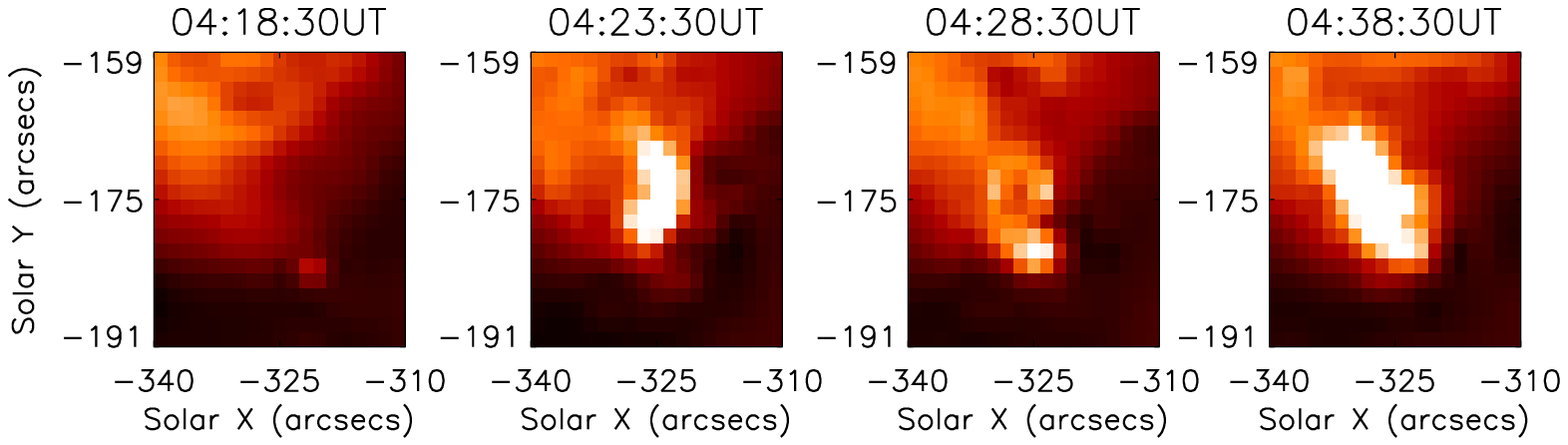}\\
\includegraphics[scale=0.70]{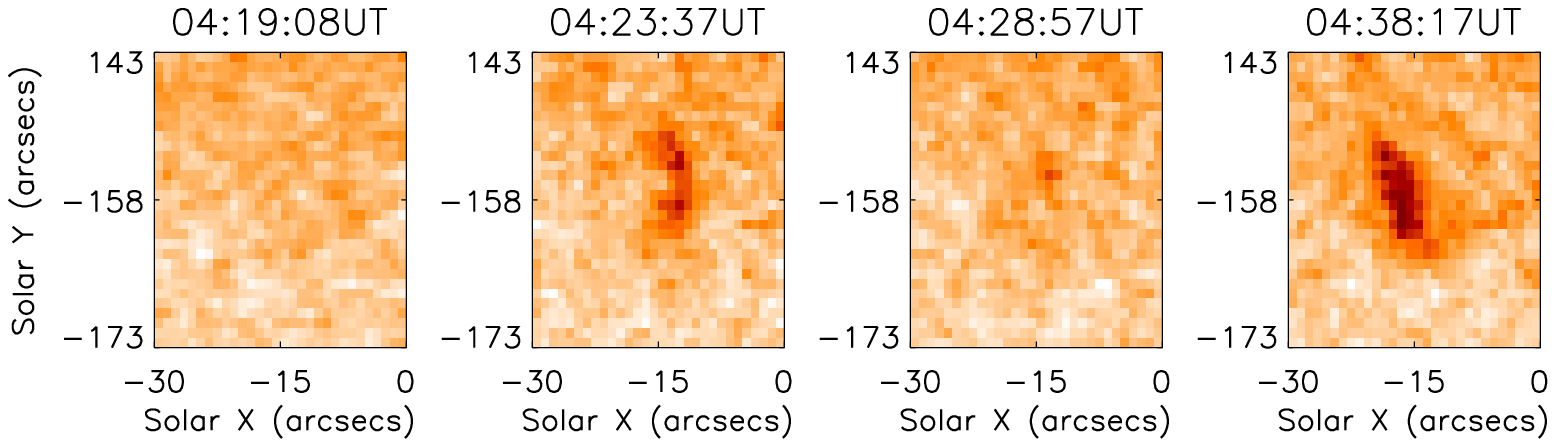}
\caption{Blinker group 2 from top  in  CDS O~{\sc v}~629~\AA, EUVI~304~\AA, 
SOT FG magnetograms, SOT Ca~{\sc ii}~H, EUVI~171~\AA\
and XRT Al$\_$poly.  The corresponding magnetogram field-of-view from which the lightcurves of 
the negative (solid line box) and positive  (dashed line) fluxes  were derived,  is outlined on the 
SOT FG magnetograms and Ca~{\sc ii}~H images.}
\label{fig4}
\end{center}
\end{figure*}

\subsection{Blinkers with X-ray counterparts}

Out of the 28 identified blinker groups, 16 showed X-ray counterparts in the XRT images with a 
lifetime of around 40 minutes in CDS O~{\sc v}~629~\AA. These blinkers were further classified 
into two categories, those associated with pre-existing X-ray coronal bright points and those 
that were not. A sample of events from each case are discussed below. 

The blinker group 1 (BG1) consisting of 15 automatically identified brightenings in CDS 
O~{\sc v}~629~\AA\ is shown in Fig.~\ref{fig2}. This figure also shows the blinker group 
as seen in EUVI~304~\AA~ (row 2), the corresponding SOT FG magnetograms in Na~{\sc i}~5896~\AA~(row 3), 
SOT Ca~{\sc ii} H (row 4), EUVI~171~\AA~ (row 5), and X-ray Al$\_$poly (row 6). Light-curves of BG1 (Fig.~\ref{fig3}) 
in CDS O~{\sc v}~629~\AA\ (thick black line), 
EUVI~171~\AA\ (thin black line), SOT Ca~{\sc ii}~H (thin grey line) and XRT Al$\_$poly (thick grey line) along with the light curve of the associated positive magnetic flux (dotted line) from the SOT magnetograms 
clearly show simultaneous intensity enhancements in all wavelengths during the brightening event. As the EUVI~304~\AA\ dataset 
has a low temporal resolution of one image every 10 minutes, we 
could not use them to follow the structure and dynamics of blinker groups over their lifetime. 
In Fig.~\ref{fig2}, BG1 can be seen as an unresolved brightening with an elongated shape 
in CDS O~{\sc v}~629~\AA\ and EUVI~304~\AA\ with a lifetime of 45 minutes in CDS O~{\sc v}~629~\AA. They show 
a sudden brightening with a faint ejection of plasma, i.e. a coronal jet, in EUVI~171~\AA\ and XRT Al$\_$poly 
data with a lifetime of 20 and 16 minutes, respectively. It is  also observed as a chromospheric 
brightening in the SOT Ca~{\sc ii}~H images with a lifetime of 52 minutes. 
%%%%%%%%% Fig 5%%%%%%%%%%%%
\begin{figure}[htp!]
\begin{center}
\includegraphics[scale=1.1]{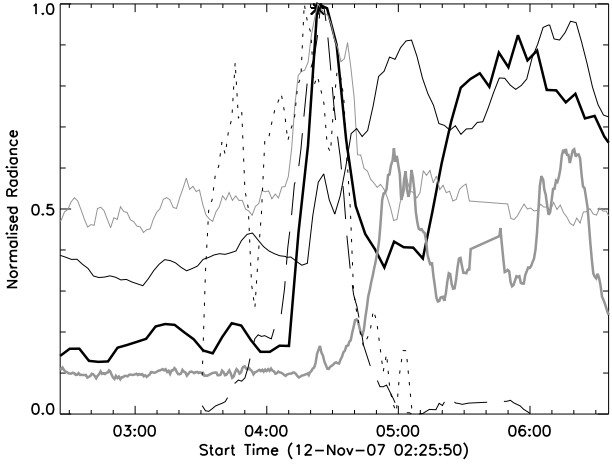}
\caption{Light-curves of blinker group 2  in CDS O~{\sc v}~629~\AA\ (thick black line), 
EUVI~171~\AA\ (thin black line), SOT Ca~{\sc ii}~H (thin grey line) and XRT Al$\_$poly (thick grey line)
along with the light-curve of the corresponding magnetic fluxes. The light curves of the positive  and the negative  magnetic fluxes in absolute value are shown with  dotted and dashed line, respectively. The asterisk symbol represents the peak in CDS O~{\sc v}~629~\AA.}
\label{fig5}
\end{center}
\end{figure}

The corresponding 
magnetogram FOV is marked with a white box in Fig.~\ref{fig2}, rows 3 \& 4. In the magnetograms, we observed a cancellation of weak bipolar fluxes prior to the event and 
an emergence of a new positive polarity flux during the course of the event. The light-curve of this newly emerged positive polarity (Fig.~\ref{fig3}) is very similar to the light-curves of the event in other wavelengths which suggests that the event was triggered by flux emergence. The brightening disappeared 
when the plasma ejection ceased as seen in the EUVI~171~\AA\ and X-ray images. This blinker/jet falls under the 
category of an event with no pre-existing features at coronal temperatures, as discussed in 
\citet{2010A&A...516A..50S}. This type of jet does not have a repetitive nature and neither 
do the corresponding blinkers. 

Another blinker group, referred to as blinker group 2 (BG2), is discussed below. Figure~\ref{fig4} 
shows O~{\sc v}~629~\AA\ data (row 1) along with EUVI~304~\AA\ images (row 2), SOT FG magnetograms 
(row 3), SOT~Ca~{\sc ii} H (row 4), EUVI~171~\AA\ (row 5) and XRT Al$\_$poly images (row 6), 
taken over the course of the blinker event. The magnetic field configuration associated 
with this event is marked with a white box (Fig.~\ref{fig4}, rows 3 \& 4). Figure~\ref{fig5} 
 shows the corresponding light-curves which have similar intensity enhancements in Ca~{\sc ii}~H 
(thin grey line), CDS O~{\sc v}~629~\AA\ (thick black line), EUVI~171~\AA\ (thin black line) and XRT 
Al$\_$poly (thick grey line) with a lifetime of 50 minutes, 43 minutes, 12 minutes and 8 minutes 
in the respective wavelengths. Figure~\ref{fig5} also shows the light-curves of the associated
positive (dotted line) and negative (dashed line) magnetic fluxes (Fig.~\ref{fig4}, row 3). 
%%%%%%%%% Fig 6%%%%%%%%%%%%
\begin{figure}[hpt!]
\begin{center}
\includegraphics[scale=0.65]{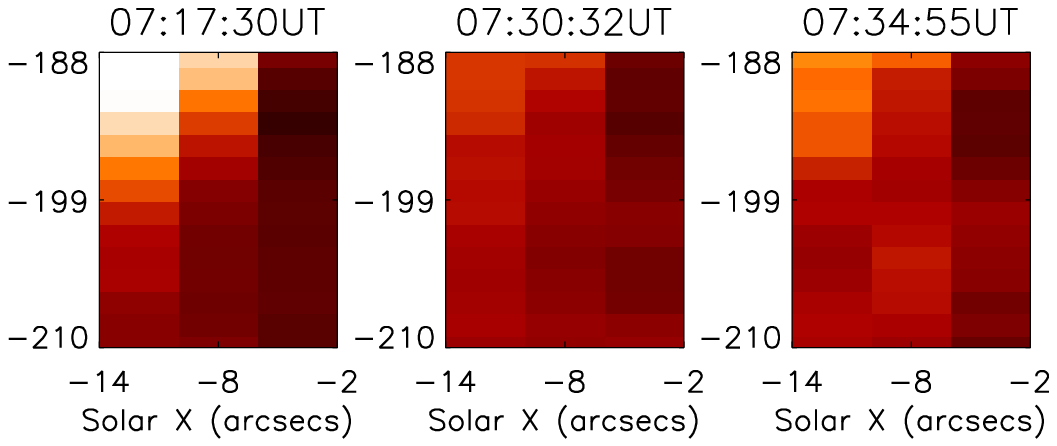}
\includegraphics[scale=0.65]{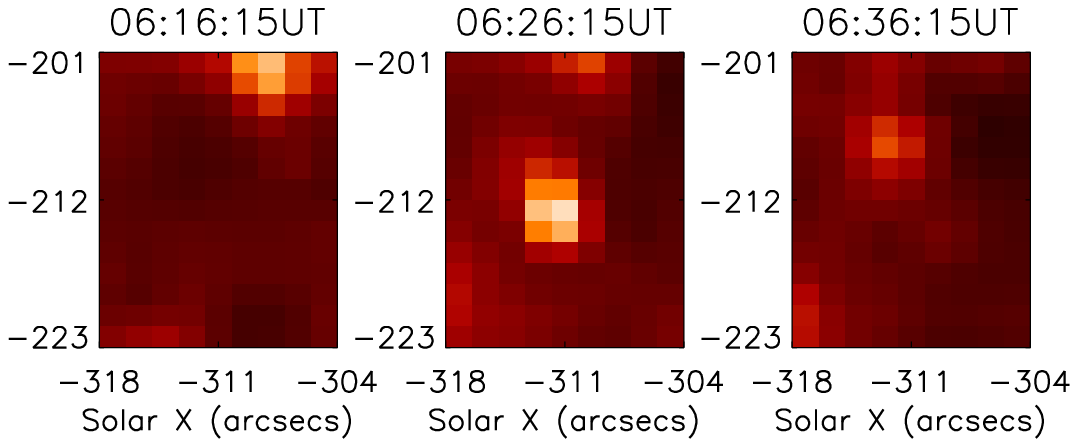}
\includegraphics[scale=0.65]{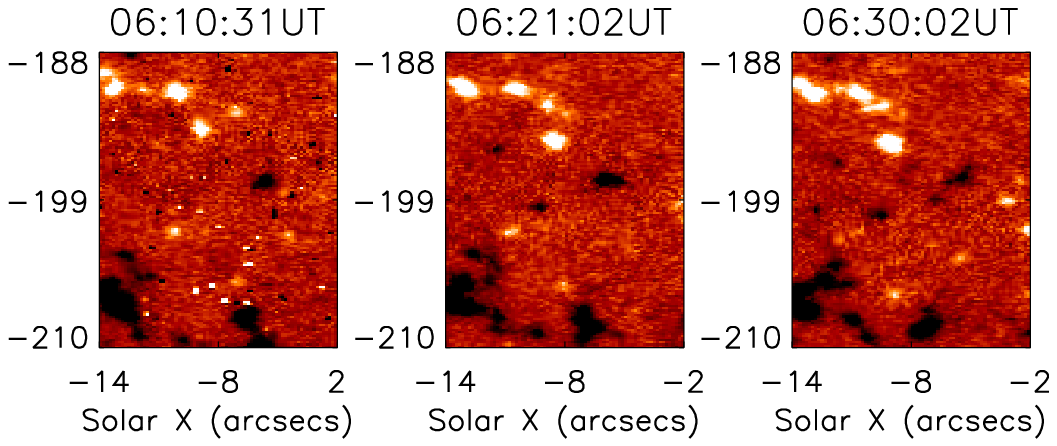}
\includegraphics[scale=0.65]{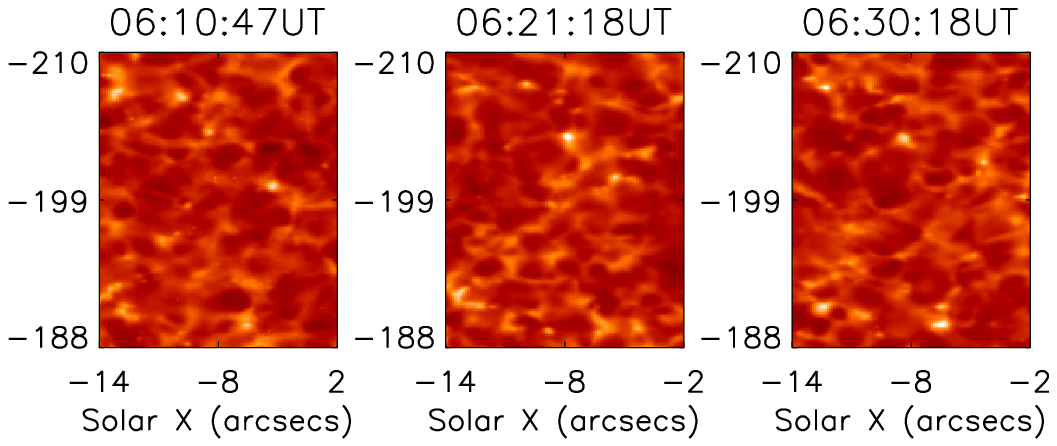}
\includegraphics[scale=0.65]{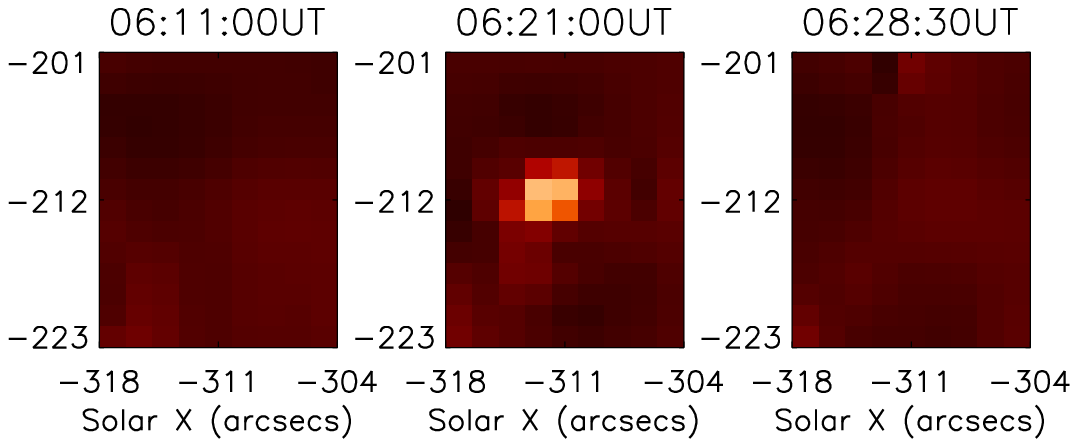}
\includegraphics[scale=0.65]{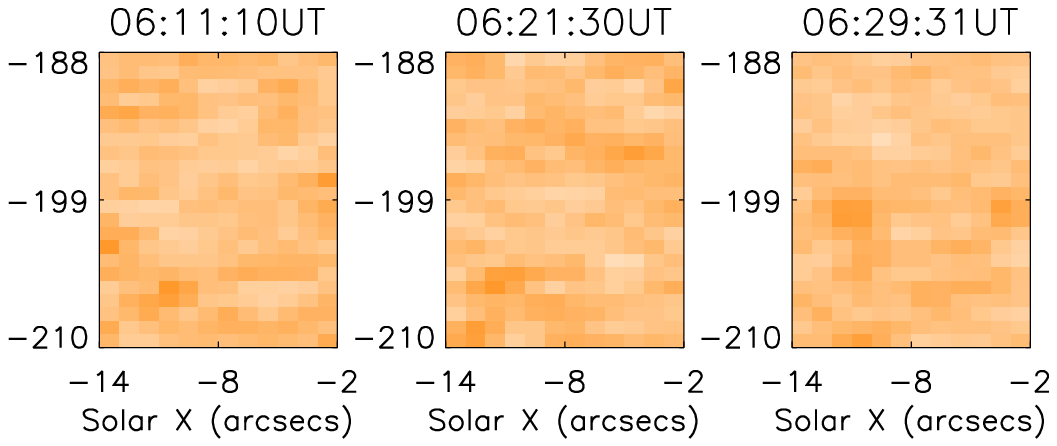}
\caption{ From top, blinker group 3 in CDS O~{\sc v}~629~\AA , EUVI~304~\AA,  SOT FG magnetograms, SOT Ca~{\sc ii}~H, EUVI~171~\AA\ and XRT Al$\_$poly.}
\label{fig6}%chbr_br2_all}
\end{center}
\end{figure}

The blinker activity started with the emergence of bipolar magnetic fluxes in the internetwork 
region as observed in the magnetograms. The peak intensity occurred around 04:23 UT in CDS O~{\sc v}~629~\AA. 
During the course of the event the pre-existing negative polarity 
flux along with the emergence of weak positive polarity fluxes (Fig.~\ref{fig4}, row 3, 
white box), resulted in a complete cancellation of weak positive polarities. SOT Ca {\sc ii} H images 
(Fig~\ref{fig4}, row 4) clearly show point-like (within the spatial resolution of the 
instrument, which is close to 100~km) 
brightening activity at chromospheric heights which corresponds to the strong negative polarity 
flux. The event is also observed at coronal heights in the EUVI~171~\AA\ (lower corona) and 
in the X-ray Al$\_$poly images as seen in Fig.~\ref{fig4} (rows 5 \& 6). The X-ray images show that 
loop structures were formed around the peak of the blinker activity. No plasma ejection were 
observed in this event at coronal heights. We can only speculate that the cancellation of the 
bipolar fluxes triggered the brightening of a pre-existing loop structure. The loop structure 
disappeared with the blinker event. 

Though we see BG2 prominently in the EUVI and X-ray data, it was not identified by the identification 
procedure. As BG2 was followed by the formation of a coronal bright point, the corresponding light 
curves were dominated by this bright point (Fig.~\ref{fig5}) at coronal temperatures (EUVI 171~\AA\ 
\& X-rays). This  caused an estimation of a very high local background and corresponding brightening 
identification threshold. Consequently, the BG2  which has a weaker radiance with respect to  the bright 
point's radiance but a clear presence at coronal temperatures, was not identified by the automatic 
procedure. This example  shows the limitations of our automated identification procedure which needs 
to be coupled with additional visual inspection in cases such as BG2.

During BG2 another positive polarity 
flux emerged (Fig.~\ref{fig4}, row 3, dotted box) which moved away from the negative polarity 
flux discussed above (Fig.~\ref{fig4}, row 3, solid box), resulting in the formation of a 
coronal bright point. Once the coronal bright point was formed at TR temperatures, it rose 
quickly to higher coronal temperatures (X-rays), producing a coronal jet in this region at around 
05:15~UT with a corresponding blinker activity. The bright point produced two more coronal 
jets/blinkers  over the course of our observations. The bright point also showed another 
6 blinker events with no coronal jets. As coronal bright points can have a range of lifetimes 
from 8 to 50 hours or longer  \citep{1974ApJ...189L..93G, 2004A&A...418..313U}, they are
likely to produce several  jets and blinkers over their lifetime. This blinker/jet falls under
the category of events associated with a coronal bright point detected at X-ray temperatures 
\citep{2010A&A...516A..50S}.  
    
\subsection{Blinkers with no X-ray counterparts}

%%%%%%%%% Fig 7%%%%%%%%%%%%

\begin{figure}[hpt!]
\begin{center}
\includegraphics[scale=1.]{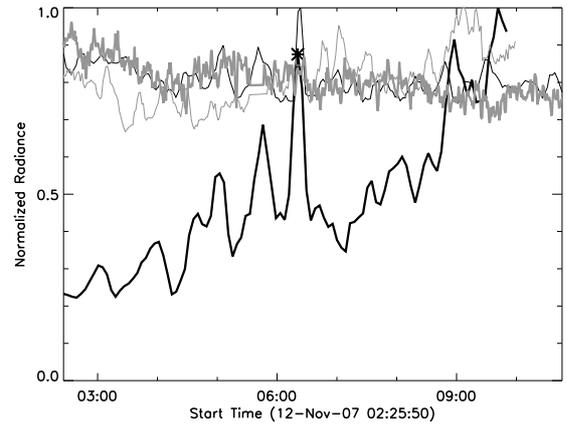}
\caption{Light-curves during blinker group 3 in CDS O~{\sc v}~629~\AA\ (thick black line), EUVI~171~\AA\ 
(thin black line), SOT Ca~{\sc ii}~H (thin grey line) and XRT Al$\_$poly (thick grey line). The asterisk 
symbol indicates the peak in CDS O~{\sc v}~629~\AA.}
\label{fig7}
\end{center}
\end{figure}

The twelve remaining blinker groups, with a lifetime of around 20 minutes in CDS 
O~{\sc v}~629~\AA, showed no detectable counterparts in the XRT images. One such group is 
discussed below, referred to as blinker group 3 (BG3).

Figure~\ref{fig6} shows BG3, consisting of two identified brightenings observed in 
CDS O~{\sc v}~629~\AA\ (row 1), EUVI~304~\AA~(row 2) and EUVI~171~\AA\ (row 5), while the 
XRT Al$\_$poly (row 6) showed no corresponding activity. The SOT FG magnetograms do not show any  
prominent changes of the magnetic fluxes (row 3) over the course of the event and no 
corresponding brightening could be detected in the Ca~{\sc ii} H data (row 4). This strongly suggests 
that the energy deposition occurred at TR heights.  Figure~\ref{fig7} also shows the light-curves of BG3 in  CDS O~{\sc v}~629~\AA\ 
(thick black line), EUVI~171~\AA\ (thin black line), SOT Ca~{\sc ii}~H (thin grey line) and XRT Al$\_$poly (thick grey line). The light-curves can be seen to peak 
simultaneously in CDS O~{\sc v}~629~\AA\ and EUVI~171~\AA. The lifetime of BG3 is 26 minutes in CDS O~{\sc v}~629~\AA, while it is 13 minutes 
in EUVI 171~\AA. This type of blinker group shows a brightening with no visible plasma ejection  at coronal heights.

\subsection{Blinkers in EIS}
\label{Blinkers in EIS}

From the above results, the question remains as to whether the observed EUVI and X-ray blinker 
counterparts are real coronal counterparts of blinkers or not. Both the EUVI~171~\AA\ 
and the XRT Al$\_$poly filters have a wide temperature response including TR temperatures (around 
logT = 5.5~K). This motivated 
us to investigate blinkers further using spectroscopic data from Hinode/EIS which is currently 
the best spectrometer for studying coronal features. We could not study the brightening 
events discussed above with EIS, as the corresponding CDS data did not have an overlap with 
the EIS field-of-view. Hence, we used only the large raster from EIS which overlaps with the CDS large raster. 
Our study has clearly indicated so far  that any blinker found 
in CDS O~{\sc v}~629.62~\AA\ has a counterpart in EUVI 171~\AA. Therefore, brightening events identified in the coronal hole
in EUVI~171~\AA\ data were used (Fig.~\ref{fig8}) to study blinker counterparts in the 
EIS observations. 
Figure~\ref{fig8} shows the intensity maps of the large EIS raster along with the 
EUVI~171~\AA\ and XRT image of the same region. We detected 22 brightening events in the EUVI~171~\AA\ images that 
correspond to the field-of-view and timing of the large EIS raster. Out of the 22 brightening events, EIS rastered 
12 of them over the course of the raster and all 12 events were clearly observed in the EIS O~{\sc v}~192.90~\AA\ 
rasters. Similar to SDO/AIA 171~\AA~filter studied recently by Del Zanna et al. (2011, accepted), the EUVI 171~\AA\ 
filter has strong contribution from TR emission which is coming from O~{\sc v} and {\sc vi} lines along with 
Fe~{\sc ix} which is sensitive to very cool, down to logT = 5.5~K emission. Single Gaussian line fit was applied 
to the EIS data to derive the intensity map, except for the
O~{\sc v}~192.90~\AA\ line which is blended in the blue wing by an Fe~{\sc xi} line. The two lines 
were separated by applying a double Gaussian line fit. Comparison of the CDS O~{\sc v}~629~\AA\ and EIS O~{\sc v}~192.90~\AA~ 
rasters (Fig.~\ref{fig9}) clearly shows that we successfully subtracted the Fe~{\sc xi} 
contribution. 

The coronal hole is indistinguishable in the O~{\sc v}~192.90~\AA\ line (logT$_{max}$ = 5.4~K), while the 
Fe~{\sc viii}~185.21~\AA\ line (logT$_{max}$ = 5.8~K)  shows a trace of the coronal hole. The coronal hole 
is clearly visible in the Fe~{\sc x}~184.590~\AA\ and Fe~{\sc xii}~195.12~\AA\ coronal lines. The 
He~{\sc ii}~256~\AA\ line is the spectral line with the lowest formation temperature available in the EIS spectral range. 
However, the interpretation of this line is complicated by blends with Si~{\sc x}~256.37~\AA, 
Fe~{\sc xiii}~256.42~\AA~ and Fe~{\sc xii}~256.41~\AA. 
%%%%%%%%% Fig 8%%%%%%%%%%%%

\begin{figure*}[hpt!]
\begin{center}
\includegraphics[scale=0.77]{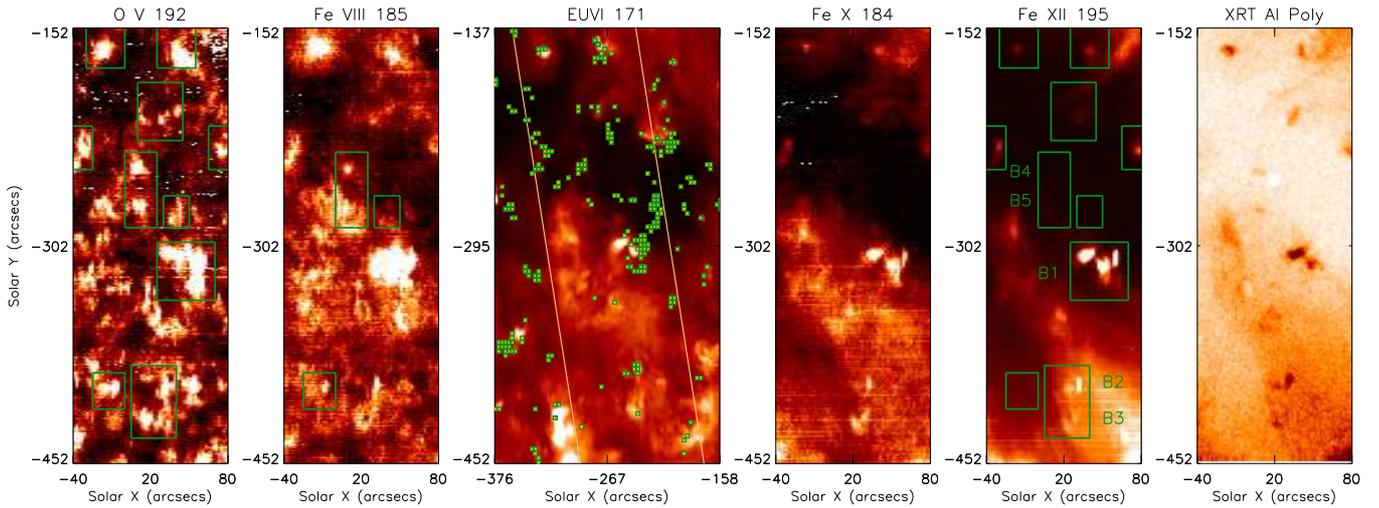}
\caption{EIS rasters of an ECH region from 2007 November 12. From left, EIS
O~{\sc v}~192.90~\AA\ and Fe~{\sc viii}~185.21~\AA\ intensity plots, EUVI 171~\AA\ image over-plotted 
with the brightenings identified as green boxes, EIS Fe~{\sc x}~184.590~\AA\ and Fe~{\sc xii}~195.12~\AA\  
intensity plots, XRT Al$\_$poly image. Box symbols over-plotted on the O~{\sc v}~629~\AA\ and Fe~{\sc xii}~195.12~\AA\ 
rasters show the 12 EUVI 171~\AA\ brightening events, with and without coronal counterparts 
at Fe~{\sc xii} temperatures. The boxes over-plotted on the Fe~{\sc viii}~185.21 raster mark events 
with no coronal counterpart in Fe~{\sc xii}~195.12.}
\label{fig8}%chbr_eis_ch}
\end{center}
\end{figure*}

\subsection{Blinkers with EIS Fe~{\sc xii} counterparts}

Out of the 12 EUVI 171~\AA\ brightenings observed by EIS, 4 of them showed counterparts in 
the EIS Fe~{\sc viii}~185.21~\AA, Fe~{\sc x}~184.590~\AA\ and Fe~{\sc xii}~195.12~\AA\ lines and were associated 
with an X-ray jet. An example event, B1, is discussed below. Figure~\ref{fig10} (left) shows a 
sequence of XRT Al$\_$poly and EUVI 171~\AA\ images during the event. The images show a coronal 
bright point and a coronal jet along with the disappearance of part of the bright point. This 
bright point persisted throughout the 10 hours of XRT Al$\_$poly observations, producing  multiple 
coronal jets. \comment{Though the EIS Fe~{\sc xii}~195.12~\AA\ intensity map shows the existence of 
a coronal bright point, the expanding loop structures were not visible which is probably 
due to strong background emission from the quiet Sun.}

Figure~\ref{fig10} (bottom) shows the light-curves of the blinker event B1 in XRT and 
EUVI 171~\AA. This event falls under the category of a blinker associated with an X-ray 
coronal bright point jet.

% \begin{figure}[h]
% \begin{center}
% \tiny\textbf{XRT Al$\_$poly}\\
% \includegraphics[scale=0.6]{plots/chbr_b5_xrt_1.eps}
% \tiny\textbf{EUVI 171 \AA}\\
% \includegraphics[scale=0.6]{plots/chbr_b5_a171_1.eps}
% \includegraphics[scale=0.9]{plots/chbr_b2eis_lc.eps}
% \vspace{-0.1in}
% \caption{ Brightening B2, with associated coronal jet and no pre-existing coronal feature 
% (Fig.~\ref{chbr_eis_ch}) showing plasma ejection as observed in movies based on the XRT Al$\_$poly data and EUVI 
% 171~\AA\ images. The bottom panel shows the light-curves of the jet, peaking simultaneously, in X-rays 
% (thick line) and in EUVI Fe~171~\AA\ (thin line).}
% \vspace{-0.25in}
% \label{chbr_b5}
% \end{center}
% \end{figure}

The brightening B2, which is marked in Fig.~\ref{fig8}, is another example of a blinker 
event with an X-ray jet. Fig.~\ref{fig11} (right) shows the event as seen in X-rays and EUVI 171~\AA. 
Figure~\ref{fig11} (bottom) shows the light-curves of the jet peaking simultaneously in X-rays 
(thick line) and  EUVI Fe~171~\AA\ (thin line). This event falls under the category of blinkers 
with an X-ray jet, but lacking a pre-existing coronal bright point in X-rays. 

Four more events showed counterparts in EIS Fe~{\sc viii}~185.21~\AA, Fe~{\sc x}~184.590~\AA\ and Fe~{\sc xii}~195.12~\AA\
with no observable plasma ejection in X-rays. An example event, B3, is discussed below. Plasma 
ejection associated with the blinker event/X-ray jet B2 (discussed above) triggered brightenings 
in the nearby loop structures giving rise to B3, as seen in the EUVI~171~\AA\ images, which 
then rose to higher coronal temperatures (log T = 6.0 and above). The blinker event B3 as seen 
in the EIS O~{\sc v}~192.90 \AA\ is probably the brightening of TR loops  and/or their footpoint 
brightening. This event belongs to the category of blinkers associated with unresolved X-ray 
brightenings as discussed by \citet{2010A&A...516A..50S}. 

\subsection{Blinkers with no EIS Fe~{\sc xii} counterparts}
%%%%%%%%% Fig 9%%%%%%%%%%%%
\begin{figure}[h]
\begin{center}
\vspace{-0.15in}
\includegraphics[scale=0.6]{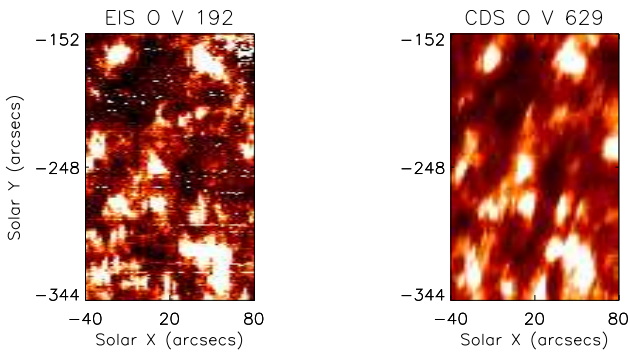}
\vspace{-0.1in}
\caption{ From left: EIS large raster in O~{\sc v}~192.90~\AA\ having common a FOV with the CDS 
large raster in O~{\sc v}~629~\AA\ taken approximately at the same time.}
\label{fig9}%chbr_cds_eis_ov}
\end{center}
\end{figure}

%%%%%%%%% Fig 10%%%%%%%%%%%%
\begin{figure}[htp!]
\begin{center}

\includegraphics[scale=0.56]{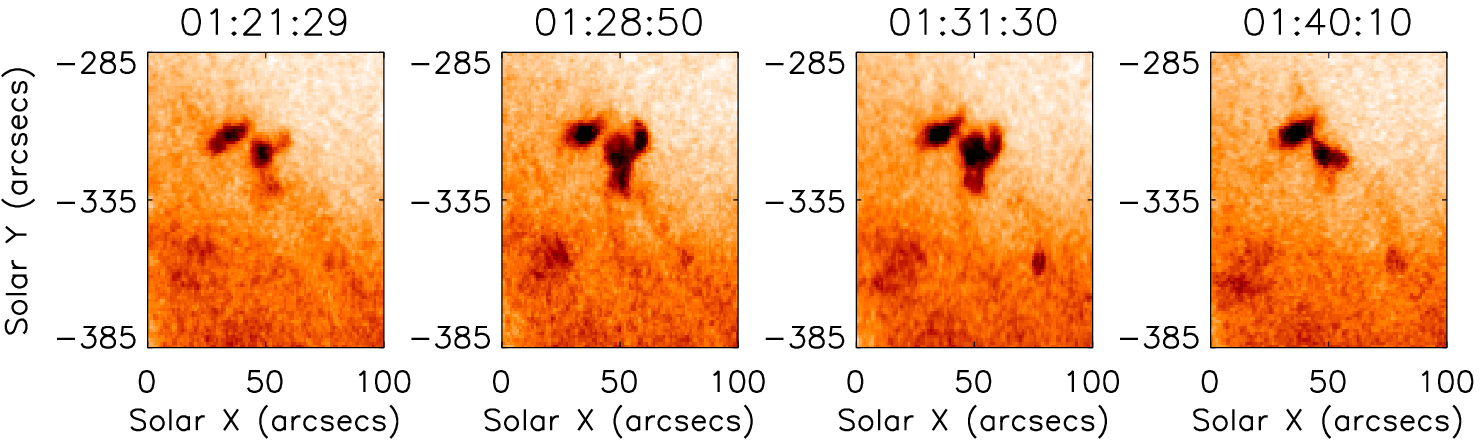}
\includegraphics[scale=0.56]{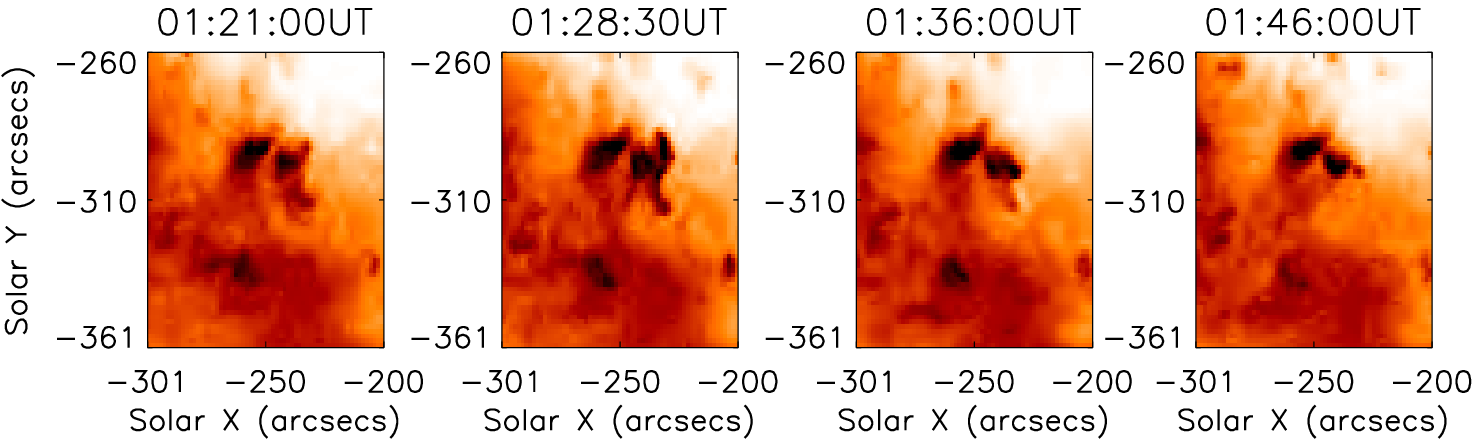}
\includegraphics[scale=0.9]{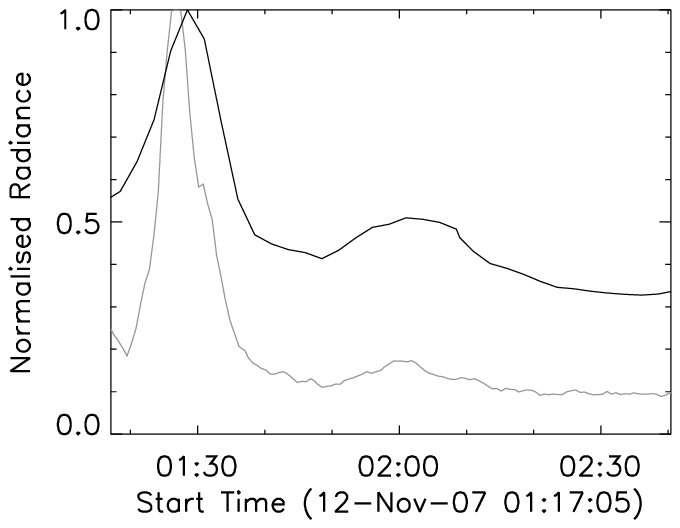}
\caption{ The coronal bright point jet B1 Fig.~\ref{fig8}) showing expansion and 
eruption of loop structures as observed in XRT Al$\_$poly and EUVI 171~\AA\ 
images. The bottom panel shows the light-curves of the blinker event B1 peaking simultaneously in XRT (grey line) and 
EUVI 171~\AA~(black line). }
\label{fig10}%chbr_b35
\end{center}
\end{figure}

%%%%%%%%% Fig 11%%%%%%%%%%%%
\begin{figure}[htp!]
\begin{center}
\includegraphics[scale=0.56]{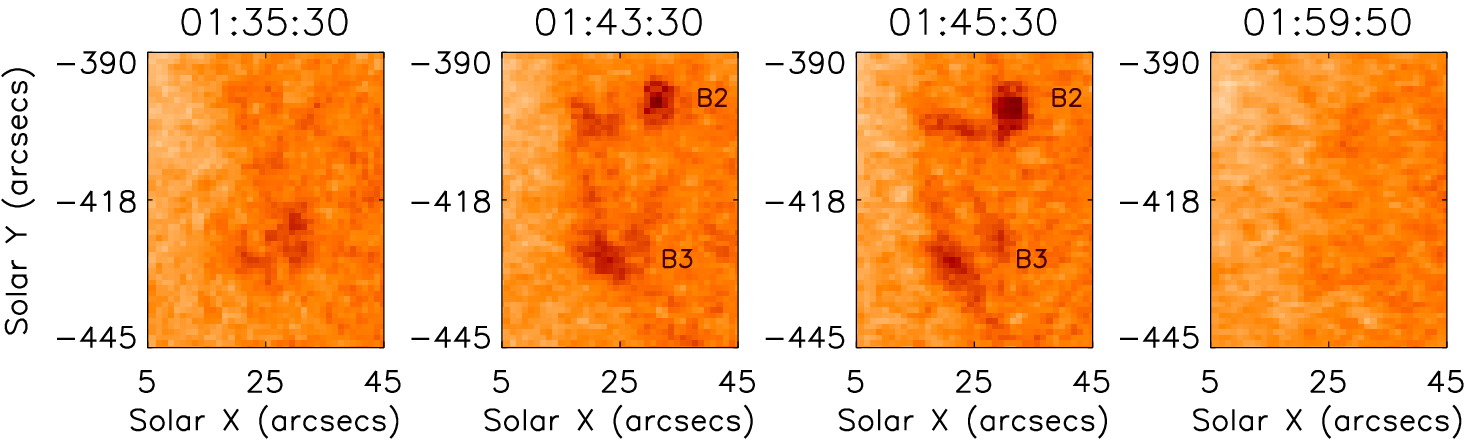}
\includegraphics[scale=0.56]{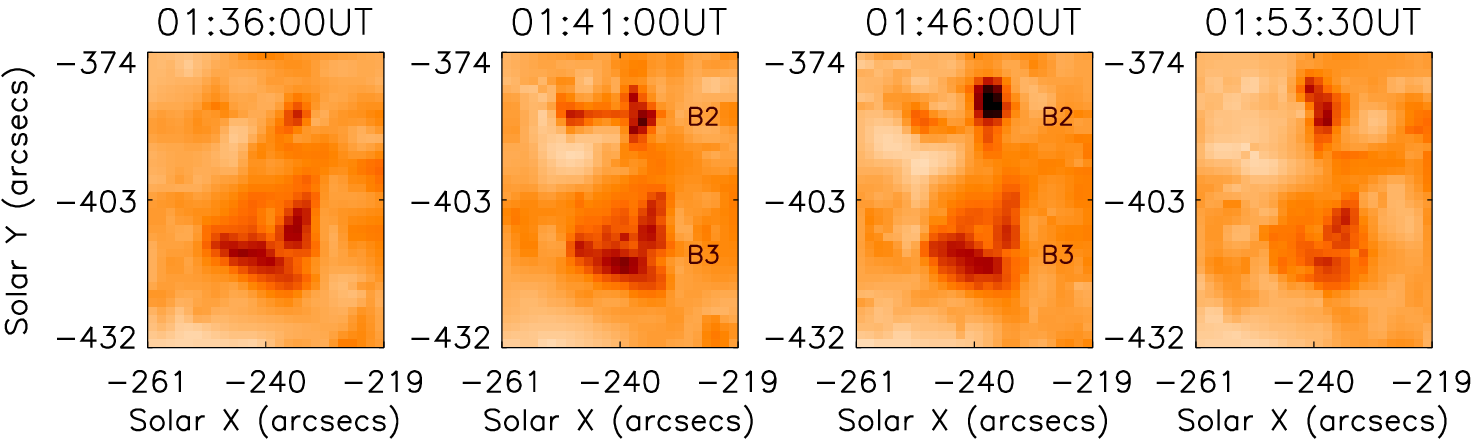}
\includegraphics[scale=0.9]{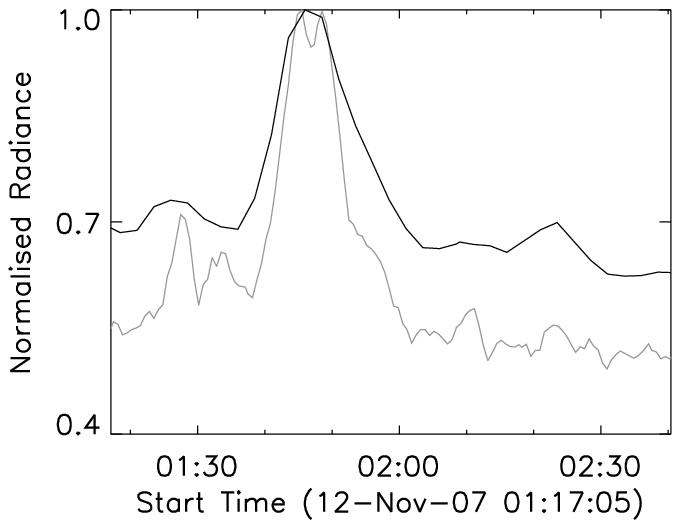}
\caption{ The  brightening B2 associated coronal jet and no pre-existing coronal feature 
(Fig.~\ref{fig8}) showing plasma ejection  as observed in XRT Al$\_$poly and EUVI 171~\AA\ 
images. The bottom panel shows the light-curves of the blinker event B2 peaking simultaneously in XRT (grey line) and 
EUVI 171~\AA~(black line). }
\label{fig11} %chbr_b35}
\end{center}
\end{figure}

The remaining 4 events showed no counterpart to the blinker in the EIS Fe~{\sc x}~184.590~\AA\ and 
Fe~{\sc xii}~195.12~{\AA} lines. Although one event was identified in X-rays, the remaining 3 events did 
not have a counterpart in X-rays. Two events showed corresponding point-like brightening in the EIS Fe~{\sc vii}~185.21~\AA\ intensity maps  (see Fig.~\ref{fig8}), while the other two showed only a dispersed brightening in that area.

%%%%%%%%% Fig 11%%%%%%%%%%%%
\begin{figure}[htp!]
\begin{center}
\includegraphics[scale=0.56]{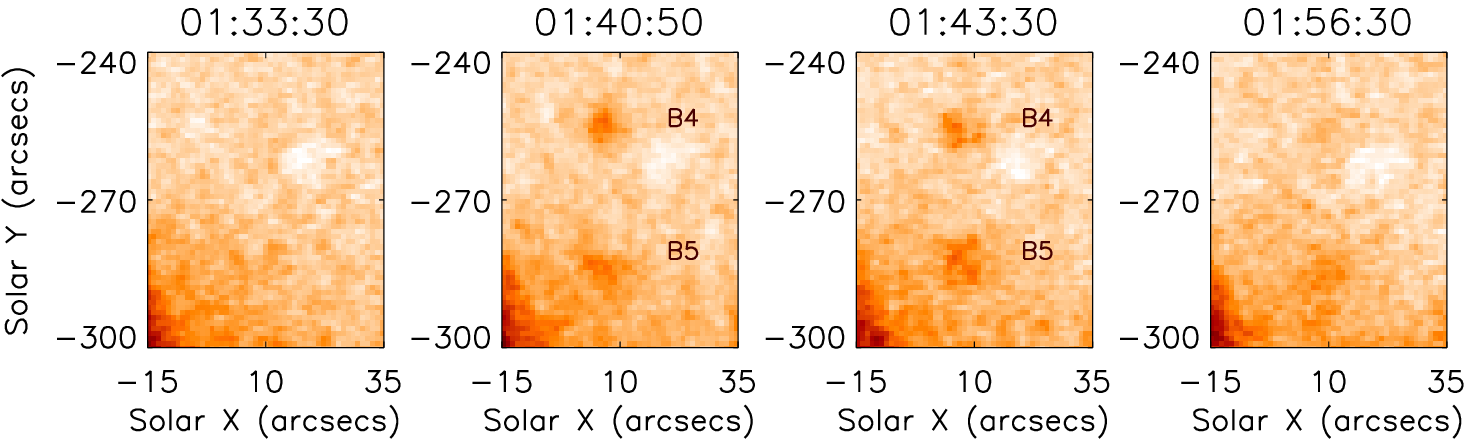}
\includegraphics[scale=0.56]{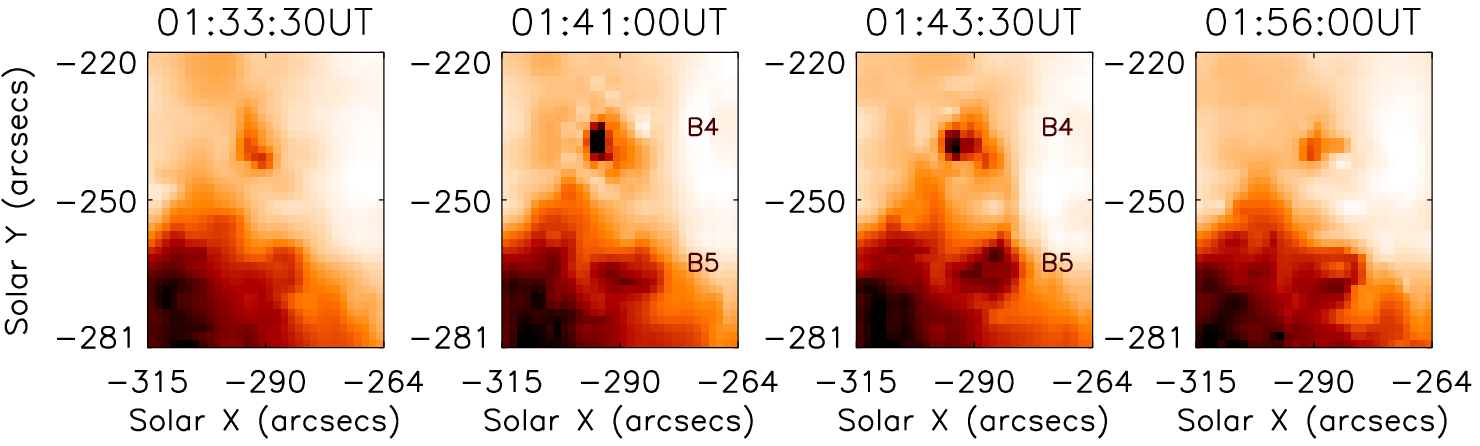}
\includegraphics[scale=0.9]{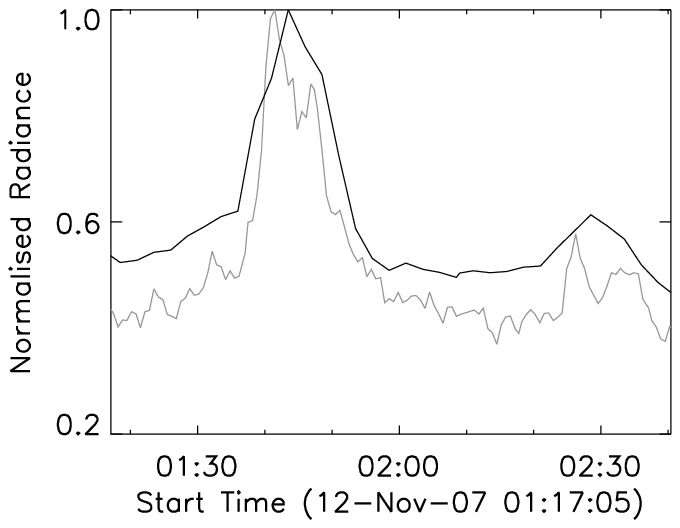}
\caption{ Brightening events B4 and B5 (Fig.~\ref{fig8}) with no coronal counterparts 
in Fe~{\sc xii}~195.12 as observed in XRT Al$\_$poly (top) and EUVI 171~\AA~(bottom) images. The 
bottom panel shows the light-curve of the brightening in X-rays and EUVI 171~\AA.}
\label{fig12}%chbr_b2}
\end{center}
\end{figure}

A sample brightening event B4, which is marked in Fig.~\ref{fig8}, is discussed below. 
The event is clearly distinguishable in the EIS O~{\sc v}~192.90~\AA\ line. In XRT data, this event falls 
under the category of unresolved brightening with no plasma ejection. While in the EUVI~171~\AA\ images 
(Fig.~\ref{fig12}), a point-like brightening (within the spatial resolution of the instrument, 
which is a few arcseconds) 
was observed with a corresponding formation/brightening of a loop structure, giving rise to 
another brightening B5 at the other end of the observed loop structure. The brightening B5 
itself is a small  loop structure in the EUVI 171\AA\ images and it was not identified in X-rays. 
In the EIS Fe~{\sc viii}~185.21~\AA\  line, brightening B4 showed a point-like brightening (again, 
within the spatial resolution of the instrument, similar to EUVI), while B5 could 
not be identified. The B5 appearance in the EUVI 171~\AA\ pass-band is, therefore, due to 
the contribution of TR emission as discussed earlier (Sec. \ref{Blinkers in EIS}). Figure~\ref{fig12} shows the light-curve 
of the brightening B4 in X-rays and EUVI 171~\AA. 

\section{Discussion and Conclusions}

The aim of the present study is to identify the true nature of the transient EUV brightenings, called 
blinkers, or, in other words, to find what causes the observed transient brightenings. We also investigated 
whether they have coronal signature, i.e., the link between EUV and coronal transient features, thereby the 
contribution of events associated with blinkers to coronal heating.

Twenty eight blinker groups were identified in CDS O~{\sc v}~629~\AA\ raster images. All of them showed 
a counterpart in EUVI 171~\AA\ and 304~\AA\ images and 57\% of them showed counterpart in X-rays. Although, 
blinkers were seen in EUVI 171~\AA\ and X-ray images, the question on whether they can reach coronal 
temperatures remained open, due to the significant contribution of transition-region emission to the 171~\AA\ 
pass-band and X-ray Al poly filters. A set of 12 blinker groups automatically identified in EUVI~171~\AA\ 
were studied with EIS in order to determine whether these brightenings have a coronal counterpart. Out of 
the 12 blinker groups, 8 were found to have an EIS Fe~{\sc xii}~195.12~\AA\ counterpart suggesting that 
events associated with the blinkers can release  thermal energy directly in the solar corona. We found, 
in general, that blinkers are associated with various coronal events, e.g. EUV/X-ray jets, brightenings 
in small-scale loops (coronal bright points) or foot-point brightenings of larger loops. 

Brightening events having no counterparts in X-rays and/or in EIS Fe~{\sc xii}~195.12~\AA\ were also 
studied. They appear as point-like brightenings in CDS~O~{\sc v}~\AA\ and EUVI 171~\AA\ images. As they do 
not reach coronal temperatures ($\geq$ log T = 6.0~K), they certainly are associated with chromospheric 
features, e.g. larger spicules \citep{2005A&A...436L..43O,2006A&A...452L..11M, 2011A&A...532L...1M, 2010ApJ...714L...1S} 
or have transition region origin, e.g. flows in small scale TR loops \citep{2004A&A...427.1065T,2008A&A...488..323S}. 
Even these blinkers reach lower coronal heights (upto logT = 5.8~K). We found that some blinkers were 
clearly associated with newly emerging magnetic flux which caused the formation of loop structures seen 
in EUVI 171~\AA\ and X-ray images. Both the loops and the blinkers faded away as the bipolar fluxes drifted 
away from each other similar to the quiet Sun blinker event studied by \citet{2008A&A...488..323S}, thereby 
confirming this to be one of the mechanisms responsible for blinkers. 

Blinkers with coronal counterpart occur repetitively and have a lifetime of around 40 min at transition 
region temperatures, while blinkers with no coronal counterpart, i.e. with transition region/chromosphere 
origin, happen only once and have a duration of 20 min in average.  In general, lifetimes of
blinkers are different at different wavelengths, i.e. different temperatures, decreasing from the chromosphere 
to the corona.

The above results show that the term blinker is a `catch-all' term for a vast range of 
transient events of coronal, transition-region or chromospheric origin. We find that the blinkers with 
coronal counterpart ($\geq$ log T = 6.0~K) are associated with coronal activities and the blinkers with no 
counterpart above log T = 6.0~K are associated with chromospheric features. Thus, up-flows from the 
chromosphere and down-flows from the corona associated with these blinkers do contribute to the formation and 
maintenance of the temperature gradient in the transition region.

\begin{acknowledgements} The authors thank ISSI, Bern for the support of the team 
``Small-scale transient phenomena and their contribution to coronal heating''. Research at 
Armagh Observatory is grant-aided by the N.~Ireland Department of Culture, Arts and Leisure. 
We also thank STFC for support via grant ST/F001843/1. Hinode is a Japanese mission 
developed and launched by ISAS/JAXA, with NAOJ as domestic partner and NASA and STFC (UK) 
as international partners. It is operated by these agencies in co-operation with ESA and 
NSC (Norway). The STEREO project is an international consortium led by the Naval Research 
Laboratory (USA), including the Lockheed Martin Solar and Astrophysics Laboratory (USA), 
the NASA Goddard Space Flight Center (USA), the Rutherford Appleton Laboratory (UK), the 
University of Birmingham (UK), the Max-Planck-Institut f\"ur Sonnensystemforschung (Germany), 
the Centre Spatial de Li\`ege (Belgium), the Institut d'Optique Th\`eorique et Appliqu\'ee 
(France) and the Institut d'Astrophysique Spatiale (France).

\end{acknowledgements}

\bibliographystyle{aa}
\bibliography{blinker_ch_refcor2}

\end{document}